\documentclass[lettersize,journal]{IEEEtran}
\usepackage[pdftex]{graphicx}
\usepackage{float}
\usepackage{cite}
\usepackage{amsmath,amssymb,amsfonts}
\usepackage{algorithmic}
\usepackage{graphicx}
\usepackage{textcomp}
\usepackage{xcolor}
\usepackage{subfigure}
\usepackage{amssymb}
\usepackage{threeparttable}
\usepackage{graphicx}
\usepackage[numbers]{natbib}
\usepackage{multicol} 
\usepackage{algorithm}
\usepackage{flushend}
\usepackage[mathscr]{euscript}
\usepackage{bm}
\usepackage{color}
\usepackage{colortbl}
\usepackage{arydshln}
\usepackage{makecell}
\usepackage{diagbox}
\usepackage{etoolbox}
\makeatletter
\patchcmd{\@makecaption}
{\scshape}
{}
{}
{}
\makeatother
\def\tablename{Table}

\begin{document}

\title{PRISE: Privacy-pReserving Image Searchable Encryption Scheme for Intelligent Vehicle Systems }

\author{{Xuemei~Fu,
	 Laurence T. Yang, ~\IEEEmembership{Fellow,~IEEE,}
	 Na~Song,
	 Jinxiong~Gao,	
     and   Zhixing~Lu}
    %

    \IEEEcompsocitemizethanks{
    	
    	\IEEEcompsocthanksitem This work was partially supported by the Joint Funds of the National Natural Science Foundation of China (No.U23A20300), the Major Science and Technology Plan Project of Hainan Province (No.ZDKJ2021044) and the Innovation Project of Hainan Province Graduate Education ( NO.Qhys2023-101) (Corresponding author: Laurence T. Yang)    	
        \IEEEcompsocthanksitem  Xuemei Fu and Jinxiong Gao are with the School of Information and Communication Engineering, Hainan University, Haikou 570100, China (e-mail: 15931012973@163.com; 21110810000007@hainanu.edu.cn).
        \IEEEcompsocthanksitem  Laurence T. Yang is with the School of Computer Science and Artificial Intelligence, Zhengzhou University, Zhengzhou 450001, China, and also with the Department of Computer Science, St. Francis Xavier University, Antigonish,  NS B2G 2W5, Canada (e-mail: ltyang@ieee.org).
        \IEEEcompsocthanksitem Na Song is with School of Mechanical, Electrical and Information Engineering, Putian University, Putian 350001, China, and also with the School of Computer and Artificial Intelligence, Zhengzhou University, Zhengzhou 450001, China(e-mail: nasong1010@163.com.)
          \IEEEcompsocthanksitem  Zhixing Lu is with the Faculty of Computer Science and Information Technology, Universiti Putra Malaysia, Serdang, 43400, Malaysia (e-mail: luzhixingo1@163.com)

    }
}



\maketitle

\begin{abstract}
With the continuous advancement of intelligent vehicle technology, the image data generated by vehicles has become increasingly critical in various applications, including driver assistance, traffic monitoring, and safety warning systems. However, this growing reliance on image data also raises pressing concerns regarding its security and privacy protection. Searchable encryption technology, as an effective means to protect data security, shows significant potential for application in the field of intelligent vehicles. In this paper, we propose a novel Privacy-pReserving Image Searchable Encryption Scheme (PRISE) to address the security and privacy challenges associated with image data in intelligent vehicles. The PRISE scheme employs Multilinear Principal Component Analysis (MPCA) to extract data features and integrates symmetric encryption and matrix encryption techniques to ensure image privacy protection. To enhance search efficiency, we utilize a Mahalanobis distance-based fuzzy C-means (FCM) clustering method, which accelerates the search process on the cloud server. We conducted comprehensive experiments to evaluate our proposed scheme, and the results were consistent with our analytical findings, confirming the security of our approach. Comparative experiments with existing schemes demonstrated that our proposed method achieves higher accuracy while maintaining a comparable query time.
\end{abstract}


\section{Introduction}
\IEEEPARstart{W}{ith} the rapid advancement of smart vehicle technology, vehicles are generating vast amounts of image data that serve critical roles in a variety of areas, including driving assistance, traffic monitoring, and safety warnings. However, the utilisation of this data also raises significant concerns regarding its security and privacy protection. As intelligent vehicles become more interconnected and reliant on data exchange, ensuring the confidentiality and integrity of image data becomes paramount. In intelligent vehicle system, image data can be visual images captured by cameras, distance and speed information obtained by radar, and three-dimensional environmental point cloud data provided by LiDAR (Light Detection and Ranging) \cite{link:41}. These different modalities of information are combined to provide a more comprehensive and accurate perception and understanding of the environment, assisting autonomous driving systems in making smarter and safer decisions.

Image data is a fundamental component of intelligent vehicle systems \cite{link:38}, owing to its abundant visual information. Relying exclusively on images from a single sensor modality may constrain the vehicle's capacity to thoroughly comprehend the environment, particularly in tough settings such as inclement weather, low-light situations, or obstructions. Consequently, there is an increasing interest in the advancement of image search methodologies designed for autonomous cars. The objective of image search for intelligent cars is to utilize data from several sensor modalities to improve perception abilities. By integrating data from several sensors, such as visible light, infrared, and radar, the vehicle can achieve a more thorough comprehension of its environment \cite{link:41}, resulting in enhanced decision-making and navigation abilities. This integration allows the vehicle to identify and recognize things, monitor their movements, and navigate securely across intricate settings.

Cloud computing is a model that enables users to access a communal array of computing resources, including servers, storage, applications, and on-demand services, via the internet \cite{link:01}. It offers numerous advantages, such as cost efficiency, adaptability, scalability, and dependability \cite{link:02,link:03}.  Cloud computing has a wide range of applications and plays a crucial role in areas such as smart cities \cite{link:43,link:45,link:46}, social networks\cite{link:44}, and recommendation systems\cite{link:47}. With the advancement of image technology, including smartphones, smart wearables, and desktop and laptop computers, the accessibility and importance of photographs, videos, and various multimedia formats have significantly expanded. These multimedia files require more storage capacity than textual materials. Consequently, cloud storage for photos offers a versatile, safe, and accessible method for organizing and collaborating on image collections \cite{link:04}. Its scalability, ease, and data safety attributes render it an advantageous option for individuals, photographers, and enterprises pursuing effective image management and collaboration solutions. Li et al. made the initial effort to concurrently facilitate data sharing and retrieval from extensive encrypted data in vehicle networks by presenting an effective multi-receiver data authorization method \cite{link:36}. Aiman et al. introduced an innovative homomorphic technique for safeguarding the privacy of image-based autonomous vehicles linked to the cloud \cite{link:37}. To facilitate navigation retrieval while maintaining destination confidentiality, \cite{link:38} introduced a secure and privacy-preserving intelligent vehicle navigation system.

The increasing reliance on cloud-based image storage solutions has prompted serious concerns about the security and privacy of sensitive visual data. Traditional encryption solutions provide data secrecy but limit the ability to do effective content-based searches on encrypted photos. The aforementioned constraint has fueled the development of image-based searchable encryption (SE) systems, which aim to balance search capabilities with data security. Image-based SE is a cryptographic technology that allows users to securely search for specified material within encrypted images without revealing the image content or compromising security \cite{link:05}. Lu et al. \cite{link:06} investigated three distance-preserving randomization strategies for encrypting features of the image in order to retain approximate resemblance between the original image and the encrypted image. Yuan et al. \cite{link:07} designed a robust K-means structure that integrates a privacy-preserving technique for comparing Euclidean distances. Wang et al. presented the first content-based image retrieval (CBIR) framework in the field of encrypted image search in order to reduce the overhead in terms of computation and transmission \cite{link:08}. Li et al. presented a CBIR technique designed to preserve confidentiality, supporting both multi-key and multi-user settings, and this scheme can resist known-background attacks \cite{link:09}. Li et al. employ convolutional neural networks (CNN) for feature extraction to enable similarity search in the context of encrypted images \cite{link:10}. However, designing an efficient and secure image-based searchable encryption scheme is challenging, as it requires balancing the trade-off between data confidentiality and search functionality. 

Existing methods for encrypted comparison of image features predominantly utilize Euclidean or Hamming distances. However, these approaches assume independence among feature dimensions and overlook the correlations between feature components, leading to reduced retrieval accuracy and increased false positives/negatives. This limitation is particularly pronounced in practical applications such as medical image classification and industrial defect detection. In contrast, Mahalanobis distance, which is based on the covariance matrix, effectively captures the correlations among feature components, eliminates the influence of scale differences, and more accurately reflects the similarity of multidimensional features. Therefore, it serves as a critical tool for enhancing the accuracy and robustness of complex image retrieval tasks.

In this work, we design a privacy-preserving encrypted image retrieval scheme based on the existing matrix encryption techniques \cite{link:11}. We introduce a fuzzy C-means clustering scheme based on Mahalanobis distance to improve the retrieval efficiency. The following are the particular contributions presented in this paper:

\begin{itemize}
	\item Multilinear principal component analysis is utilized to derive the feature matrix from image tensors, maintaining picture compression quality while significantly decreasing storage requirements.
	\item Symmetric encryption and matrix encryption are utilized to protect the secrecy of image tensors. It has the capacity to reduce the storage overhead of cloud servers and the computational load on data owners
	\item We use the fuzzy C-means method to categorize encrypted image tensors based on Mahalanobis distance, which improves image retrieval efficiency and helps cloud servers search more efficiently.
\end{itemize}

{\bfseries Organization.} The following sections are arranged as follows: an overview of the body of knowledge and current research in the topic is provided in Section ~\ref{sec:rel}. The system model, threat assumptions, privacy requirements and preliminaries are presented in Section~\ref{sec:pro}. The comprehensive construction methodology is outlined in Section~\ref{sec:overv}. Subsequently, a comprehensive security analysis is presented in Section~\ref{sec:sec}, followed by a thorough examination of performance and evaluation in Section~\ref{sec:perf} . In conclusion, the findings of our study are presented in Section~\ref{sec:fut}.

\begin{table*}
	\tiny
	\centering
\caption{Comparison between prior works and our scheme}
	\centering
	\begin{threeparttable}
		\resizebox{\textwidth}{12mm}{
			\begin{tabular*}{\hsize}{@{}@{\extracolsep{\fill}}cccccc@{}}
				\hline
				& $Data \ type^{1}$ & $Feature \ type^{2}$ & $Three - party ^{3} $&  $Encrypted \ feature \ comparison ^{4}$ &$ Query \ privacy^{5}$ \\
				\hline
				$\lbrack26 \rbrack$ & Text & Keywords & $\times$  &  $\times$ & $\checkmark$ \\
				$\lbrack31\rbrack $ & Image & Keywords & $\times$ &  $\times$& $\checkmark$ \\
				$\lbrack32\rbrack $ & Image & Part \ of \ the \ image\ feature  & $\times$ &  $\times$  &$\times$ \\
				$\lbrack33\rbrack $ & Image & Feature \ vector  & $\checkmark$  & $\checkmark$  & $\checkmark$ \\
				
				$\lbrack35\rbrack $ & Image & Feature \ vector &$\checkmark$ &  $\checkmark$ &$\checkmark$   \\
				PRISE &  Multimodal data  & Tensor feature &  $\checkmark$ & $\checkmark$ &$\checkmark $  \\

				\hline
		\end{tabular*}}
		\begin{tablenotes} 
			\footnotesize   
			\item[1]Data type represents the type of data processed by the scheme. Text type refers to unstructured data. The image type represents data composed of pixel points. The multimodal data encompasses various types of data, including text, image, and audio data.
			\item [2]Feature type refers to the key characteristics used to represent data during data encryption and search.
			\item [3]Three-party refers to whether the scheme supports the search result of the SE scheme in which the three parties, that is, the data owner, server, and user. 
			\item [4]Encrypted feature comparison refers to the process of comparing features without disclosing sensitive information, such as keyword, distance information.
			\item [5]Query privacy refers to the protection of the privacy of the query data during the process of searching over encrypted data, meaning that the content of the query is not disclosed to unauthorized third parties.
			\end{tablenotes}  	
		\label{table:compare}
	\end{threeparttable} 
\end{table*}

\section{Related Work}
\label{sec:rel}
{\bfseries Image search and applications.} Image search is a prerequisite for content-based detection, which can identify and classify specific objects in images. Image detection can be applied to many scenarios, such as autonomous driving, security monitoring, biometric recognition, road condition detection, social networks, etc. A cloud-based image coding scheme was introduced by Yue et al., which diverges from existing image coding schemes in that it reconstructs images from a large-scale image database using descriptions rather than pixel-by-pixel compression \cite{link:24}. The work in \cite{link:25}, \cite{link:26} illustrates how to use a sizable image database to retrieve, compose, and rebuild a single image. With the use of web images, an innovative approach to image denoising was introduced by Yue et al., which incorporates both internal and external correlations \cite{link:27}. Martin-Brualla et al. presented a method for generating time-lapse videos of popular locations by combining a large number of highly linked photos available online \cite{link:28}. Dong et al. proposed two high confidence and large-scale near repeat image detection techniques based on entropy filtering and graph cutting. It preserves high-quality features while simultaneously improving picture search performance \cite{link:29}.

{\bfseries Encrypted storage outsourcing.} The utilization of cloud service providers enables clients with limited resources to outsource substantial volumes of data effectively and at a lower cost \cite{link:12}. The proliferation of massive amounts of image data has grown faster, outsourcing images to the cloud, due to its abundant computational resources, has become a popular trend. There is a notable issue that has arisen regarding the protection of confidential information while allowing for the use of external picture providers. To address this challenge, Wang et al. introduced a new architecture for an outsourced image recovery service, which offers a unique approach to recovering images \cite{link:13}. As images become increasingly important in people’s daily lives, CBIR has been the subject of considerable scholarly investigation. In the context of cloud storage, plaintext-domain CBIR techniques are not applicable. To address this challenge, a scheme that enables CBIR on encrypted images while preventing the disclosure of sensitive private data to the cloud server was proposed by Xia et al. \cite{link:14}. Hu et al. introduced a method that leverages cloud computing to deliver compressive sensing reconstruction and image storage services, all while ensuring the confidentiality and effective management of the images \cite{link:15}.

{\bfseries Secure search over encrypted data.} Song et al. first introduced a (SE) scheme for encrypted data, which enables untrusted cloud servers to remain oblivious to any information about the plaintext \cite{link:16}. The majority of these methods pertain to the retrieval of textual data through the use of keywords \cite{link:17, link:18, link:19}. There exist a limited number of systems designed for the purpose of privacy preservation in content-based image retrieval. The initial privacy-preserving CBIR method was introduced by Lu et al. \cite{link:20}, which relied on order-preserving encryption and min-hash techniques. Yuan et al. utilized a Bag-of-Words model to extract features from images. They preserved the privacy of images by encrypting the features, and subsequently performed similarity retrieval on the encrypted high-dimensional features \cite{link:21}. Huang et al. proposed a privacy-preserving CBIR scheme that uses relevance feedback to improve the retrieval accuracy \cite{link:22}. The research suggests an effective approach called Secure Similar Image Matching (SESIM) to ensure the confidentiality of outsourced image features. This protocol operates in an encrypted domain. The computational burden of the put-forward SESIM protocol is being contrasted with that of established secure distance measures \cite{link:23}. Tong et al. proposed a verifiable fine-grained encrypted image retrieval scheme \cite{link:39}. Zhu et al. proposed a medical image retrieval scheme achieving accuracy and privacy preservation through encrypted data \cite{link:40}.

A systematic comparative analysis was conducted with other searchable encryption schemes. The assessment is conducted by comparing various features related to secure image retrieval, such as data types, feature types, support for multi-party computations, encrypted feature comparison, query privacy, and scalability. From Table~\ref{table:compare}, it can be concluded that the proposed PRISE scheme outperforms existing solutions.

\section{Problem Formulation}
\label{sec:pro}

In this section, we formalize the system overview, threat assumptions, privacy requirements, and preliminaries.
\subsection{ System Model}
\begin{figure}[!ht]
	\centering
	\includegraphics[width=3.5in]{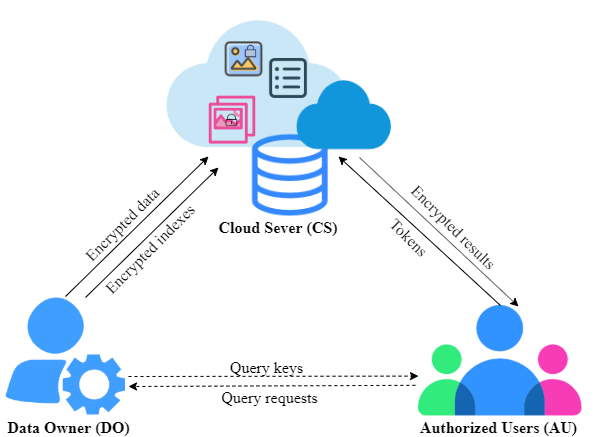}
	\caption{System model for PRISE}
	\label{fig:systemmodel}
\end{figure}

Our work focuses on analyzing a cloud-based image feature search system that encompasses three distinct parties: data owner (DO), authorized users (AU), and cloud server (CS) (as shown in Fig.\ref{fig:systemmodel}). The DO stores image data in the cloud, while AU perform search operations on the data stored on the cloud server. Our main focus on how to use cloud-based images search services without compromising privacy. In intelligent vehicle systems, DO refers to the vehicles provider, while AU are the customers who purchase the vehicles. In this context, images refer to the data collected by various sensors, such as cameras, radar, etc.

\begin{itemize}
	\item {\bfseries DO:} DO serves as the owner of image data, constructing an index by extracting features from pictures. Finally, the encrypted images and secure indexes are transferred to the CS. The owner of the image also provides users with authorization to access the data stored on the CS.
	\item {\bfseries AU:} AU is a data user authorized by DO, who conducts queries on data stored on the cloud server by submitting image query tokens. The cloud server utilizes these tokens to perform similarity searches on the data, subsequently returning query results to the user.
	\item {\bfseries CS:} CS offers data storage services for the DO, effectively managing issues related to insufficient local storage and computational resources. Additionally, the cloud server offers query services to authorized users.
	
\end{itemize}

\subsection{ Threat Assumptions}
We assume that the DO and AU are trustworthy in our system model. The DO stores encrypted image data on an honest-but-curious cloud server and authorizes AU to submit correct query tokens to the cloud server for executing queries.

Based on the knowledge possessed by potential attackers or cryptanalytic adversaries, we consider the Known-Plaintext Attack (KPA) model, and define its attack capability as follows:\\

{\bfseries Definition 1.} \emph{KPA}. Building upon the foundation of Ciphertext-Only Attack (COA), it is presumed that the attacker has the ability to acquire a collection of tuples comprising queries, images, and indices. Furthermore, the attacker is assumed to have knowledge of the ciphertext associated with these tuples.

\begin{enumerate}
	\item[1).] {\bfseries DO:} As illustrated in Fig.\ref{fig:systemmodel}, the described system model consists of three separate entities: the DO, AU, and CS. We assume that the DO is trustworthy and stores the encrypted data on the CS. Additionally, the DO provides query keys to authorized users, allowing them to access the data. Before uploading the image data to the cloud server, the DO manages the database and encrypts it. The DO also creates a secure index for the encrypted image database. It is important to note that the DO is responsible not only for preserving the security of the data but also for ensuring that the data remains easily searchable by other users.
	
	\item[2).] {\bfseries AU:} We assume that the user is trustworthy. Before gaining access to data stored on the cloud, the user must first acquire permission from the DO. For data users, the retrieval process is simplified. They only need to submit a query token. Upon receiving the query token, the CS will execute the query operation and return the results to the user.
	
	\item[3).] {\bfseries CS:} We make the assumption that the CS is honest-but-curious, a stance that aligns with the majority of prior research \cite{link:18, link:19}. This means that the CS will strictly adhere to the protocol for storing image data and providing retrieval services to authorized AUs. However, due to its curiosity about the image data, the CS may attempt to infer and analyze the data, whether it is encrypted images or encrypted vectors.
	
\end{enumerate}

In our model, we assume that the adversary has complete knowledge of the encryption scheme, excluding the key. In real-world scenarios, the adversary may have access to varying levels of background information. Accordingly, we carefully define threat models tailored to various scenarios.

\emph{Attack -\uppercase\expandafter{\romannumeral1}}: The cloud server is assumed to be semi-trusted and treated as a potential adversary. The server has access only to the encrypted image database $D$ and all encrypted images feature $\tau$ used for search queries. This model aligns with the ciphertext-only attack (COA) model, which is commonly employed in the security evaluation of data encryption protocols \cite{link:11}.

In \emph{Attack -\uppercase\expandafter{\romannumeral1}}, while the adversary is limited to accessing encrypted data, their analytical capabilities can still pose significant threats to the security of the system. The system's privacy protection heavily depends on the robustness of its encryption mechanisms. To mitigate the risks associated with the COA model, searchable encryption systems must employ more sophisticated encryption techniques and enhanced privacy-preserving mechanisms, ensuring that plaintext information remains secure even if the adversary gains access to the encrypted data.

\emph{Attack-\uppercase\expandafter{\romannumeral2}}: Based on \emph{Attack Scenario 1}, we assume that the adversary possesses some plaintext image samples but lacks knowledge of their corresponding encrypted data. This represents a known-sample attack as described in [13]. For instance, an attacker may observe an encrypted image dataset while obtaining plaintext data samples from other sources. This allows them to know the content of certain images in the plaintext database. In the context of autonomous driving, the adversary could be the vehicle owner who has access to some plaintext images collected by the car’s sensors. The data owner, in this case, is the vehicle manufacturer, who processes the data to enhance the car’s performance.

 In \emph{Attack-\uppercase\expandafter{\romannumeral2}}, the cloud server is considered a semi-trusted adversary with access to the encrypted image database and all encrypted images used for search queries. Additionally, the adversary may possess some plaintext image samples but lacks knowledge of their corresponding encrypted representations. By comparing the known plaintext samples with the encrypted data, the adversary could attempt to infer patterns or weaknesses in the encryption scheme. This poses a potential risk to system security, particularly if vulnerabilities exist in the encryption algorithm. Therefore, when designing searchable encryption schemes, it is essential to account for such attack scenarios and implement targeted measures to enhance both system security and privacy protection.

\emph{Attack -\uppercase\expandafter{\romannumeral3}}: Based on \emph{Attack Scenario 1}, we consider the possibility of collusion between the cloud server and the user. In this case, the adversary, in addition to accessing the encrypted image dataset, could encrypt images of interest to the user and perform encrypted queries. Such attacks are particularly relevant in certain real-world applications. For instance, the cloud server could handle image queries submitted by the user, thereby potentially revealing the content of the user's candidate images.

In \emph{Attack -\uppercase\expandafter{\romannumeral3}}, the cloud server, through collaboration with the user, can encrypt images of interest to the user and perform encrypted queries. This may reveal the user's search intentions, preferences, or sensitive information, leading to a breach of user privacy. If the encryption scheme contains vulnerabilities, the cloud server might exploit additional information obtained through its collaboration with the user (eg., encrypted query images) to analyze patterns in the encryption mechanism, potentially inferring the content of unauthorized images in the database. Therefore, when designing searchable encryption systems, it is crucial to implement rigorous mechanisms for protecting data and query privacy to prevent potential information leakage.

\subsection{Privacy Requirements}
To guarantee data privacy for each entity under the previously mentioned framework and threat model, it is necessary to address the following privacy considerations concurrently.
\begin{enumerate}
	\item[1)]\emph{Images security.} Images are the private property of DO and may contain sensitive information related to the data owner. Encrypted image data is stored on the cloud server, and image security ensures that only authorized users can access and decrypt the data. Simultaneously, the image data remains protected against unauthorized access or tampering by the cloud server.
	
	\item[2)]\emph{Index confidentiality.} In searchable encryption, indices are typically constructed to efficiently store information about encrypted data, facilitating rapid searches. In the context of encrypted image searches, the index consists of feature vectors extracted from the images. Index confidentiality, therefore, means that attackers should be unable to derive any information about the original images from the encrypted feature vectors.
	
	\item[3)]\emph{Queries confidentiality.} In searchable encryption schemes, query confidentiality ensures the secrecy of search queries. When an authorized AU submits a search query to a cloud server, the query content needs to be protected to prevent unauthorized users or attackers from gaining insights into the user's search interests. Since image searches are conducted using query feature vectors, query confidentiality means that attackers should not be able to extract any information about the original images from the query tokens or search results.
	
	\item[3)]\emph{Tokens unlikability.} When performing image searches on a cloud server, users typically generate search tokens and initiate search requests. Token unlikability ensures that these tokens are non-traceable and indistinguishable from one another, preventing any relationships between the tokens from being inferred.
	
\end{enumerate}

\begin{figure}[!ht]
	\centering
	\includegraphics[width=3.3in]{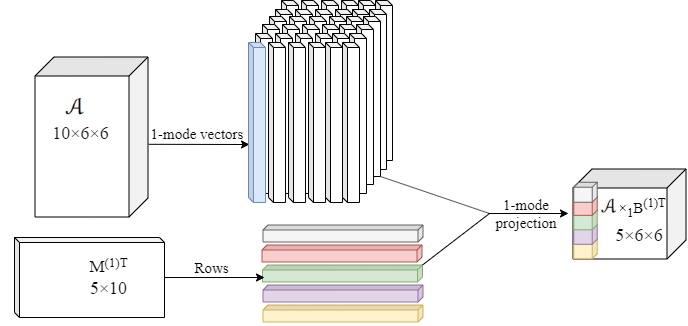}
	\caption{Multilinear projection of tensor objects.}
	\label{fig:mpca}
\end{figure}

\begin{figure*}[htbp]
	\centering
	\includegraphics[width=1.2\textwidth,angle=0,height=3.4in]{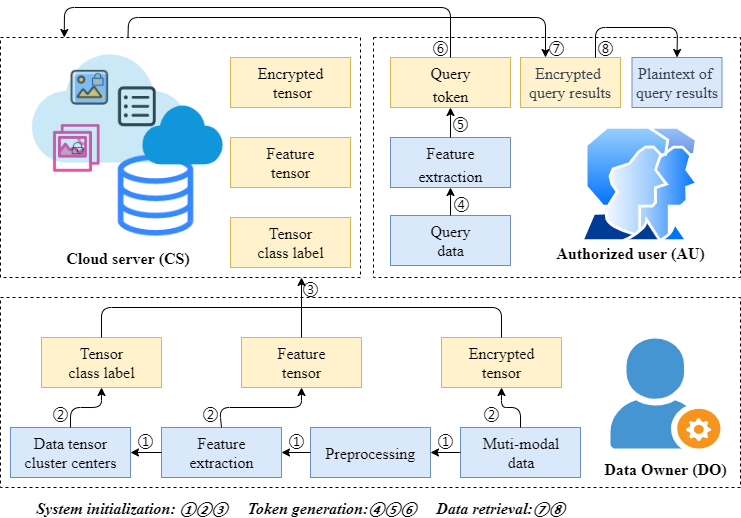}
	\caption{Overview of PRISE.}
	\label{fig:framework}
\end{figure*}

\subsection{Preliminaries}
The preliminary stages of the tensor \cite{link:30}, multilinear principal component analysis (MPCA) \cite{link:31}, and fuzzy C-means classification based on Mahalanobis distance \cite{link:32,link:33} will be discussed in this section. 

{\bfseries Tensor.} Tensors are a generalization of matrices, commonly referred to as multidimensional arrays. Formally, an N-th-ways tensor is denoted by the notation $\mathcal{T}$ $\in$ $\emph{R}^{I_1 \times I_2 \times...\times I_N}$, where $I_N$-th represents the size of the tensor along the $N$-th dimension. In the field of image processing, matrices are typically used to represent images. Each element of the matrix corresponds to a pixel in the image, and pixel values can be directly stored in the matrix elements. Consequently, the use of tensors facilitates the efficient storage and processing of image data. 

{\bfseries Multilinear algebra.} In accordance with the principles of standard multilinear algebra \cite{link:31}, it is possible to represent any tensor $\mathcal{T}$ as the product.
\begin{equation*}
\mathcal{T}=\mathcal{S}\times_{1}\emph{D}^{(1)}\times_{2}\emph{D}^{(2)}\times...\times_{N}\emph{D}^{(N)} \quad (1)
\end{equation*}

where $\mathcal{S}$=$\mathcal{T}\times_{1}$ $\emph{D}^{(1)^{T}}\times_{2}$ $\emph{D}^{(2)^{T}}\times...\times_{N}$ $\emph{D}^{(N)^{T}} $ and $\emph{D}^{n}$=\\ $(\emph{d}_{1}^{(n)}\emph{d}_{2}^{(n)}...\emph{d}_{I_{n}}^{(n)})$ is an orthogonal $I_{n}\times I_{n}$ matrix.

{\bfseries Multilinear projection of tensor objects.} 
An $N$th-order tensor $\mathcal{T}$ exists within the tensor (multilinear) space $\emph{R}^{I_1}\otimes{R}^{I_2}\otimes...\otimes{R}^{I_N}$, where ${R}^{I_1},{R}^{I_2},...,{R}^{I_N}$ are the $N$ linear vector space \cite{link:34}. For typical image and video tensors, tensors can be employed to represent information such as pixels, colors, brightness, and frames evolving over time within images and videos. By representing images and videos as tensors, tensor operations can be leveraged for tasks such as image processing, feature extraction, and object recognition. To achieve the dimensionality reduction of a tensor subspace, it is essential to identify a tensor subspace capable of capturing the predominant features of the tensor. These features can be utilized for image classification and recognition.

To achieve this goal, $Q_{n}<I_{n}$ orthogonal basis vectors for each modality's n-mode linear space are sought, forming a tensor subspace $\emph{R}^{Q_1}\otimes{R}^{Q_2}\otimes...\otimes{R}^{Q_N}$ comprised of these linear subspaces. Let $\widehat{D}$ denote a matrix containing $Q_{n}$ orthogonal n-mode basis vectors. The projection of $\mathcal{T}$ onto the tensor subspace $\emph{R}^{Q_1}\otimes{R}^{Q_2}\otimes...\otimes{R}^{Q_N}$ is defined as $\mathcal{T}^{'}= \mathcal{T}\times_{1} \widehat{D}^{(1)^{T}}\times_{2} \widehat{D}^{(2)^{T}}\times...\times_{N} \widehat{D}^{(N)^{T}}$. A visual representation of the tensor multilinear projection is shown in Fig.\ref{fig:mpca}. In Fig.\ref{fig:mpca}, A third-order tensor $\mathcal{A}$$\in$$\emph{R}^{10\times 6\times 6}$ is projected onto a one-mode vector space using a projection matrix $\emph{M}\in \emph{R}^{5\times 10}$, resulting in a projected tensor $\mathcal{A}\times_{1}\emph{M}^{(1)^{T}}$.

{\bfseries Fuzzy C-means based on Mahalanobis distance.} Fuzzy c-means (FCM) algorithm: the FCM technique is considered the most prevalent and effective among many fuzzy clustering algorithms. It accomplishes the automated categorization of sample data by optimizing the objective function to calculate the degree of membership of every point in the sample to all cluster centers, thereby determining the class membership of sample points. Each sample is assigned a membership function for each cluster, and samples are classified based on the magnitude of membership values. He et al. proposed the FCM based on Mahalanobis distance (FCM-MD) algorithm \cite{link:34}, which partitions a finite dataset into c fuzzy clusters using the MD function. Specifically, the objective function of the FCM-MD algorithm is as follows:
\begin{equation*}
J_{FCM_{MD}}=\sum_{i=1}^{c}\sum_{j=1}^{n}u_{ij}^{m}\cdot D_{MD}^{2}(x_{j},\vec{c}_{i})\quad (2)
\end{equation*}
where $m\in{[1,+\infty)}$ represents fuzzy control, $u_{ij}$ is an element of the membership matrix. and $D_{MD}$ is the Mahalanobis distance. Notably, $u_{ij}$ indicates the membership value of $x_{j}$ in the $i$-th cluster, and $\vec{c}_{i}$ denotes the cluster center of the $i$-th cluster.

\section{The Construction of PRISE}
\label{sec:overv}

In this section, we propose a privacy-preserving searchable symmetric encryption scheme based on image tensors, called PRISE. The PRISE scheme primarily involves three stages: 1) system initialization, 2) trapdoor generation, and 3) similar image retrieval and reading. An overview of PRISE scheme is depicted in Fig.\ref{fig:framework}. First, the DO performs standardization, noise-removing, feature extraction, and image tensor clustering operations on the image tensors, generating encrypted image tensor, feature tensor, and tensor class labels. The DO uploads the encrypted image tensors, feature tensors, and tensor class labels to the CS. When an authorized AU initiates a query, the AU extracts features from the queried image and generates an encrypted image token, which is sent to the CS. Finally, the CS executes a similarity query and returns the encrypted similar image tensors to the AU.

\subsection{ System Initialization}

\begin{figure}[!ht]
	\centering
	\includegraphics[width=3.2in]{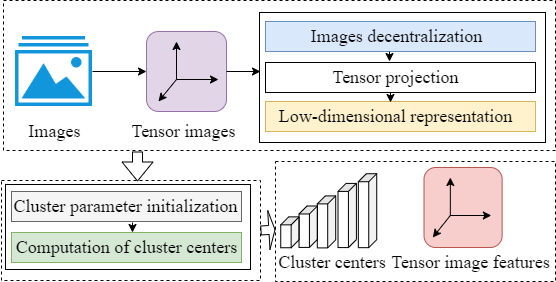}
	\caption{PRISE data preprocessing.}
	\label{fig:init}
\end{figure}

In this phase, the data owner initializes the image data, with the specific steps shown in Fig.\ref{fig:init}. The data owner first represents the data uniformly as tensors. Next, Multilinear Principal Component Analysis (MPCA) is applied to extract image features. During this step, the data undergoes decentralization, tensor projection, and dimensionality reduction, resulting in a low-dimensional image tensor. Following this, the MDFCM clustering method is employed to cluster the images. Finally, the clustering centers and the tensor image features are output. Data initialization and outsourced storage are accomplished through Algorithm \ref{alg:MPCA}for tensor images data dimensionality reduction and Algorithm \ref{alg:ENC} for data encryption.

Specific tensor dimensionality reduction steps:

\begin{itemize}
	\item \textit{{Step 1.} Centering}
	\par
	For the original image tensor $\mathcal{I}_{m}\in \mathbb{R}^{I_{1}\times I_{2}\times ...\times I_{N}},m=1,2,...,M$, center each mode (dimension) by subtracting the mean $\overline{\mathcal{I}}_{m}$ along that mode. This step is similar to centering in traditional PCA.
	\item  \textit{{Step 2.} Eigen-decomposition}
	\par
	Perform eigenvalue decomposition on the covariance matrix $\sum_{m=1}^{M}\widetilde{I}_{m(n)}$. This results in eigenvectors $\widetilde{H}^{n}$. The covariance matrix captures the relationships and variances between different modes.
	
	\item \textit{{Step 3.} Projection matrix construction}
	\par
	Choose the top-$k$ eigenvectors representing the $k$ biggest eigenvalues. These eigenvectors form the projection matrix $\widetilde{H}^{(n)}$.
	\item \textit{{Step 4.} Low-rank image tensor approximation}
	\par
	Project the centered tensor onto the subspace defined by the selected eigenvectors. This step results in a low-rank approximation $\mathcal{Y}_{m}=\mathcal{I}_{m}\times_{1}\widetilde{U}^{(1)^{T}}\times_{2}\widetilde{U}^{(2)^{T}}\dots \times_{N}\widetilde{U}^{(N)^{T}} ,m=1,2,...,M \rbrace$  of the original tensor $\mathcal{I}_{m}$.
\end{itemize}
\begin{algorithm}[H]
	\caption{Init}
	\label{alg:MPCA}
	\begin{algorithmic}[1] 
		\REQUIRE ~~\\ 
		A set of tensor images $\lbrace \mathcal{I}_{m}\in \mathbb{R}^{I_{1}\times I_{2}\times ...\times I_{N}},m=1,2,...,N  \rbrace$, declare the initial quantity of clusters $c$, the degree of fuzziness $m$, the convergence error $\epsilon$\\
		\ENSURE ~~\\ 
		Low-dimensional results $\lbrace \mathcal{Y}_{m}\in \mathbb{R}^{P_{1}\times P_{2}\times ...\times P_{N}},m=1,2,...,N \rbrace$, $\xi_{m}$, $c_{i}$
		\STATE  Images sample decentralization  $\lbrace \widetilde{\mathcal{I}}_{m}=\mathcal{I}_{m}-\overline{\mathcal{I}}_{m},m=1,2,...,N \rbrace$, where $\overline{\mathcal{I}}_{m}=\frac{1}{M}\sum_{m=1}^{M}(\overline{\mathcal{I}}_{m})$ is the sample mean
		\STATE The eigen-decomposition of $\phi^{(n)\ast}=\sum_{m=1}^{M}\widetilde{I}_{m(n)}\dot{\widetilde{I}^{T}_{m(n)}}$ and set $\widetilde{H}^{n}$ to comprise the eigenvectors corresponding to the prominent $P_{n}$ eigenvalues, where $n=1,...,n$ 
		\STATE  Calculate  $\lbrace \widetilde{\mathcal{Y}}_{m}=\widetilde{\mathcal{I}}_{m}\times_{1}\widetilde{D}^{(1)^{T}}\times_{2}\widetilde{D}^{(2)^{T}}\dots \times_{N}\widetilde{D}^{(N)^{T}} ,m=1,2,...,N \rbrace$
		\STATE Calculate $\psi_{y_{0}}=\sum_{m=1}^{N}\parallel \widetilde{\mathcal{Y}_{m}}\parallel^{2}_{F} $
		\STATE \textbf{for} $k=1:K $ \textbf{do}
		\STATE \quad \textbf{for} $ n=1:N$ \textbf{do}
		\STATE \quad \quad set the matrix $\widetilde{D}^{(n)}$ to consist of the $P_{n}$ eigenvectors of $\phi^{(n)}$
		\STATE \quad \quad Calculate$\lbrace \widetilde{\mathcal{Y}}_{m},m=1,2,...,N\rbrace$ and $\psi_{y_{k}}$
		\STATE \quad \quad if $\psi_{y_{k}}-\psi_{y_{k-1}}<\eta$
		\STATE \quad \quad \quad  break 
		\STATE  The $\mathcal{I}_{m}$ after projection is obtained as $\mathcal{Y}_{m}=\mathcal{I}_{m}\times_{1}\widetilde{D}^{(1)^{T}}\times_{2}\widetilde{D}^{(2)^{T}}\dots \times_{N}\widetilde{D}^{(N)^{T}} ,m=1,2,...,N \rbrace$ 
		\STATE Rearrange $\mathcal{Y}_{m}$ into vector $\xi_m$
		\STATE Init the membership matrix $U^{(k)}=[u_{ij}]_{c\times n}$, subject to$\sum_{i=1}^{k}u_{i,j}=1$,
		$\forall j\in{1,2,...,n}$, $\sum_{j=1}^{n}=u_{i,j}\textgreater 1,\forall i\in{1,2,...,k}$ 
		\STATE Update the centers according to $c_{i}=\frac{\sum_{j=1}^{n}u^{m}_{i,j}i_{j}}{\sum_{j=1}^{n}u^{m}_{i,j}}$,where $i=1,2,...,c$
		\STATE Update the fuzzy covariance matrix according to $\sum_{i}=\frac{\sum_{j=1}^{n}u^{m}_{i,j}(i_j-c_i)(i_j-c_i)^{T}}{\sum_{j=1}^{n}u^{m}_{i,j}}$
		
		\STATE Update the memberships values according to $\frac{1}{\sum_{l=1}^{k}\left[\frac{(i_j-c_i)^{T}\sum_{i}^{-1}(i_j-c_i)-ln|\sum_{i}^{-1}|}{(i_j-c_l)^{T}\sum_{i}^{-1}(i_j-c_l)-ln|\sum_{l}^{-1}|}\right]^{\frac{1}{m-1}}}$ and store in a matrix $U^{(k+1)}$ 
		\STATE if $\lVert U^{(k+1)}-U^{(k)}\rVert <\epsilon$ 
		
		\RETURN  $\mathcal{Y}_{m}$, $\xi_{m}$, $c_{i}$
	\end{algorithmic}
\end{algorithm}

\begin{figure}[!ht]
	\centering
	\includegraphics[width=3.2in,angle=0,height=1.4in]{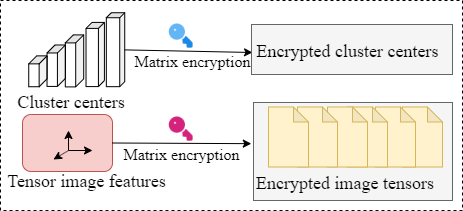}
	\caption{PRISE data encryption mechanism.}
	\label{fig:enc}
\end{figure}

\begin{itemize}
	\item \textit{{Step 5.} Cluster centers}
	\par
	First, rearrange $\mathcal{Y}_{m}$ into vector $y_m$. Then, determine cluster centers $c$ based on the FCM-MD.
	
\end{itemize}

\begin{algorithm}[H]
	\caption{DataEnc}
	\label{alg:ENC}
	\begin{algorithmic}[1] 
		\REQUIRE ~~\\ 
		Database $\xi=\lbrace \xi_{1},\xi_{2},...,\xi_{N}\rbrace$, $D$'s covariance matrix $\sum_{D}$, cluster centers and covariance
		matrices $(c_i,\sum_{i})$,1$\leq i \leq k$ , secret key $sk_1=\lbrace M_{1},M_{2} \rbrace$,$sk_2=\lbrace M_{3},M_{4} \rbrace$ \\
		\ENSURE ~~\\ 
		Encrypted database $\lbrace E(\xi_{1}),E(\xi_{2}),...E(\xi_{M})\rbrace $\\
		Encrypted cluster centers $\lbrace E(C_1),E(C_2),...,E(C_k)\rbrace$\\
		\textbf{Image encryption}
		\STATE \textbf{for} 1$\leq p \leq N$ \textbf{do}\\
		\STATE \quad $W_{p}=\xi_{p}\cdot \sum^{-1}_{D}\cdot \xi^{T}_{p}$
		\STATE \quad $\widehat{\xi_{p}}=\xi_{p}\cdot{(\sum^{-1}_{D})}^{T}$
		\STATE \quad $\overline{\xi_{p}}=\xi_{p}\cdot \sum^{-1}_{D}$
		\STATE \quad Randomly choose a number $\gamma_{i}$
		\STATE \quad $\widehat{\xi_{p}}[M]= \gamma_{p}$,$\widehat{\xi_{p}}[M+1]=1$
		\STATE \quad $\overline{\xi_{p}}[M]=W_{p}$,$\overline{\xi_{p}}[M]=-\gamma_{p}$
		\STATE \quad Initialize two $(M+2)\times(M+2)$ matrices $\widehat{I_{p}}, \overline{I_{p}} $
		\STATE \quad \textbf{for} $0\leq x\leq M+1,0\leq y\leq M+1$  \textbf{do}
		\STATE \quad \quad if $x==y$ then
		\STATE \quad \quad \quad $\widehat{I_{p}}[x,y]=\widehat{I_{p}}[x], \overline{I_{p}}[x,y]=\overline{\xi_{p}}[x]$
		\STATE \quad \quad else $\widehat{I_{p}}[x,y]=0, \overline{I_{p}}[x,y]=0$
		\STATE Randomly choose two lower triangular matrices $\widehat{Q_{p}}, \overline{Q_{p}}$ with diagonal entries set to 1
		\STATE $ E(\widehat{\xi_{p}})=M_{1}\widehat{Q_{p}} \widehat{I_{p}}M_{2}$
		\STATE  $ E(\overline{\xi_{p}})=M_{1}\overline{Q_{p}} \overline{I_{p}}M_{2}$
		\STATE $E(\xi_{p})\gets \lbrace E(\widehat{i_{p}}),E(\overline{i_{p}}) \rbrace$\\
		\textbf{Cluster centers encryption}
		\STATE  \textbf{for} 1$\leq i \leq k$  \textbf{do}
		\STATE \quad $E(c_{i})$= \textbf{Image encryption}$(c_i,\sum_{i})$ // execute lines 2 to 16 of the code
		\STATE \quad Initialize an $(M+2)\times(M+2)$ matrix $\sum^{'}_{i}$
		\STATE \quad  \textbf{for} 1$\leq i \leq M+1$,1$\leq i \leq M+1$  \textbf{do}
		\STATE \quad \quad  \textbf{if } i$<$M\&\&j$<$M 
		\STATE  \quad\quad \quad \quad $\sum^{'}_{i}[i][j]=\sum^{-1}_{i}[i][j]$
		\STATE  \quad \quad \textbf{else} $\sum^{'}_{i}[i][j]=randomnumber()$
		\STATE $E(\sum_i)=M_3\sum^{'}_{i}M_4$
		\STATE $E(C_i)\gets \lbrace E(c_{i}),E(\sum_i)\rbrace  $		
		\RETURN Encrypted database $\lbrace E(\xi_{1}), E(\xi_{2}),...,E(\xi_{M}) \rbrace$\\
		Encrypted cluster centers $\lbrace E(C_1),E(C_2),...,E(C_k)\rbrace$
	\end{algorithmic}
\end{algorithm}

The specific initialization algorithm pseudocode is illustrated in Algorithm \ref{alg:MPCA}. In the image encryption phase, DO encrypts images, extracts feature vectors, and performs clustering to obtain c cluster centers. DO stores the encrypted images and indices in CS, where CS provides image retrieval services for AU.

Data encryption primarily consists of two components: one involves encrypting image features, and the other entails encrypting cluster centers. As shown in Fig.\ref{fig:enc}, in the data encryption phase, the data owner encrypts the cluster centers and image features using matrix encryption.

Images encryption steps:
\begin{itemize}
	\item \textit{{Step 1.} Image encryption}
	\par
	Given a database $\xi=\lbrace i_{1},i_{2},...,i_{M}\rbrace$, the covariance matrix  $\sum_{D}$ of  $D$, and the key $sk_1=\lbrace M_{1},M_{2} \rbrace$, the algorithm outputs the encrypted database $\lbrace$ $E(i_{1}), E(i_{2})$,..., $E(i_{M})\rbrace$. The specific algorithm is illustrated in Algorithm \ref{alg:ENC}.
\end{itemize}

\begin{itemize}
	\item \textit{{Step 2.} Cluster centers encryption}
	\par
	To accelerate image search speed, this paper utilizes the FCM-MD clustering algorithm based on matrix encryption for image encryption. This method not only expedites search speed but also ensures the privacy protection of data, preventing unauthorized disclosure. The algorithm achieves the goal of securing data by expanding the matrix and introducing randomness. Given cluster centers and covariance matrices $(c_i,\sum_{i})$, 1$\leq i \leq k$ and secret key $sk_1=\lbrace M_{1},M_{2} \rbrace$,$sk_2=\lbrace M_{3},M_{4} \rbrace$, the algorithm executes the portion related to the encryption of cluster centers. Ultimately, the algorithm outputs the encrypted cluster centers $\lbrace E(C_1),E(C_2),...,E(C_k)\rbrace$.	
\end{itemize}	

\begin{algorithm}[H]
	\caption{TokenGen}
	\label{alg:tokengen}
	\begin{algorithmic}[1] 
		\REQUIRE ~~\\ 
		Query iamge feature $\tau=({q_{1},q_{2},...,q_{M}})$, secret key $sk=\lbrace M_{1},M_{2} \rbrace$, $\lbrace M_{3},M_{4} \rbrace$ \\
		\ENSURE ~~\\ 
		Token $Q_{\tau}$ \\
		\textbf{construct a $(M+2)\times(M+2)$ matrix}
		\STATE \textbf{ for}1$\leq j\leq n$ \textbf{do}
		\STATE \quad $\psi_{j}=q_{j}\dot({q_{1},q_{2},...,q_{M}}) $
		\STATE constructs a $M\times M$ matrix $\Psi$
		\STATE  $\Psi=[\psi_{1},\psi_{2},...,\psi_{M}]^{T}$ 
		\STATE extends $\Psi$ to $(M+2)\times(M+2)$ matrix as $\Psi^{'}$, $\Psi^{'}=$
		\begin{equation}
		\left[
		\begin{array}{cc}
		\psi & R^{1}_{M\times 2}  \\ 
		R^{2}_{2\times M} & R^{3}_{2\times 2}
		\end{array}
		\right]
		\nonumber
		\end{equation}
		where $R^{1}_{M\times 2}$,$R^{2}_{2\times M}$, $R^{3}_{2\times 2}$ are three random matrices chosen by AU.\\
		\textbf{Generate query token}
		\STATE Initialize two vectors $\widehat{\tau}=\overline{\tau}=-\tau$
		\STATE Randomly choose two number $\lambda$ and $\gamma_{\tau} $
		\STATE $\widehat{\tau}[M]=\lambda, \widehat{\tau}[M+1]=\gamma_{\tau}$, 
		\STATE $\overline{\tau}[M]=1,\overline{\tau}[M+1]=\lambda $
		\STATE Initialize two $(M+2)\times(M+2)$ matrices $\widehat{Q}, \overline{Q} $
		\STATE \textbf{for} $0\leq x\leq n+1,0\leq y\leq n+1$ \textbf{do}
		\STATE \quad \textbf{if} $x==y$ 
		\STATE  \quad \quad \quad$\widehat{Q}[x,y]=\widehat{\tau}[x], \overline{Q}[x,y]=\overline{\tau}[x]$
		\STATE \quad \textbf{else} $\widehat{Q}[x,y]=0, \overline{Q}[x,y]=0$
		\STATE Randomly choose two lower triangular matrices $\widehat{Q_{\tau}}, \overline{Q_{\tau}}$ with diagonal entries set to 1
		\STATE $ E(\widehat{\tau})=M^{-1}_{2}\widehat{Q} \widehat{Q_{\tau}}M^{-1}_{1}$
		\STATE $E(\overline{\tau})=M^{-1}_{2}\overline{Q} \overline{Q_{\tau}}M^{-1}_{1}$
		\STATE $E(\tau)\gets \lbrace E((\widehat{\tau})),E(\overline{\tau}) \rbrace$
		\STATE $E(\Psi^{'})=M^{-1}_{4}\Psi^{'}M^{-1}_{3}$
		\STATE Token $Q_{\tau}$= $\lbrace E(\tau),E(\Psi^{'}) \rbrace$
		
		\RETURN Token $Q_{\tau}$
	\end{algorithmic}
\end{algorithm}

The specific data encryption algorithm pseudocode is illustrated in Algorithm \ref{alg:ENC}.
\begin{figure}[!ht]
	\centering
	\includegraphics[width=3.2in]{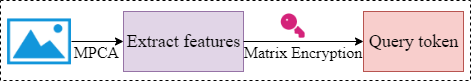}
	\caption{PRISE token generation process flow.}
	\label{fig:token}
\end{figure}

\subsection{Trapdoor Generation}

In the token generation phase, authorized users first extract the image features using Multilinear Principal Component Analysis (MPCA), and then encrypt the features using matrix encryption to generate query tokens, as shown in Fig.\ref{fig:token}. The user first expands the query vector to an $(M+2)\times(M+2)$ matrix, then encrypts the vector and matrix using keys $sk=\lbrace M_{1},M_{2} \rbrace$ and $sk=\lbrace M_{3},M_{4} \rbrace$, respectively. Ultimately, the algorithm outputs the query token $Q_{\tau}$. The specific algorithm is illustrated in Algorithm \ref{alg:tokengen}.

\subsection{ Similar image Retrieval and Reading}

\begin{algorithm}[H]
	\caption{Query}
	\label{alg:query}
	\begin{algorithmic}[1] 
		\REQUIRE ~~\\ 
		Encrypted database $\lbrace E(\xi_{1}), E(\xi_{2}),...,E(\xi_{M}) \rbrace$, $\lbrace E(C_1),E(C_2),...,E(C_k)\rbrace$, the token $E(\tau)$ \\
		\ENSURE ~~\\  A collection $R_{\tau}$ \\
		\textbf{Determine the cluster}
		\STATE \textbf{for} $1\leq i\leq c$ \textbf{do}
		\STATE \quad $ \widehat{R_{i}}=E(\widehat{C_{i}})\cdot E(\widehat{\tau_{i}}) $
		\STATE \quad $ \overline{R_{i}}=E(\overline{C_{i}})\cdot E(\overline{\tau_{i}}) $
		\STATE  \quad ${R_{i}}=E(\Psi^{'})\cdot E(\sum_i) $
		\STATE \quad $ CD_{\tau}=Tr(\widehat{R_{i}})+Tr(\overline{R_{i}})+Tr({R_{i}})$
		\STATE  CS identifies the minimum distance $CD_{\ell}$ within $CD_{\tau}$
		\textbf{Compute the top-k elements within cluster $CD_{\ell}$}
		\STATE \textbf{for} $1\leq i\leq N$ \textbf{do}
		\STATE \quad $ \widehat{P_{i}}=E(\widehat{i_{i}})\cdot E(\widehat{\tau_{i}}) $
		\STATE \quad $ \overline{P_{i}}=E(\overline{i_{i}})\cdot E(\overline{\tau_{i}}) $
		\STATE \quad $ D_{i}=Tr(\widehat{P_{i}})+Tr(\overline{P_{i}})$
		\STATE Sort $D_{i}$ in ascending order to form a new collection $R_{\tau}$
		\RETURN Result $R_{\tau}$
	\end{algorithmic}
\end{algorithm}

\begin{figure}[!ht]
	\centering
	\includegraphics[width=3.0in]{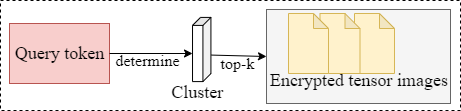}
	\caption{PRISE query mechanism .}
	\label{fig:query}
\end{figure}

In the query phase, the cloud server CS first identifies the cluster to which the query vector belongs. Subsequently, it determines the top-k results of the query vector within this cluster, as shown in Fig.\ref{fig:query}. Since all computations on the cloud server CS are performed under ciphertext, no plaintext information of the data owner DO is disclosed.

The server CS receives a query token $E(\tau)$ from the user AU, first, determines the class to which the query token $E(\tau)$ belongs through these computation $ \widehat{R_{i}}=E(\widehat{C_{i}})\cdot E(\widehat{\tau_{i}}) $, $ \overline{R_{i}}=E(\overline{C_{i}})\cdot E(\overline{\tau_{i}})$, $CD_{\tau}=Tr(\widehat{R_{i}})+Tr(\overline{R_{i}})+Tr({R_{i}})$. Based on these values $\lbrace CD(C_1),CD(C_2),...,CD(C_c)\rbrace$, identify the minimum value $CD_{\ell}$. Subsequently, the cloud server CS computes the distance between the query vector $E(\tau)$ and each vector in this $CD_{\ell}$ class. The distance set $\lbrace D_{\ell 1}, D_{\ell 2},..., D_{\ell n} \rbrace$ is then arranged in descending order, and the encrypted images corresponding to the top-k results are selected as the user's query results, which are returned to the AU.

\section{Security Analysis}
\label{sec:sec}

This section primarily evaluates the security of the PRISE scheme. The evaluation focuses on two aspects: correctness and security. The correctness of the scheme is established through formula derivation, while the security is assessed using the real/ideal simulation approach.

\subsection{Correctness Analysis of PRISE}
We validate the correctness of the scheme through formula derivation and demonstrate that the new probabilistic encryption algorithm does not compromise the accuracy of the final results. The correctness analysis of this scheme is deduced following the reference \cite{link:11}. From the above algorithms, it is evident that the correctness of the PRISE scheme relies on the formula of $D_{i}\le D_{j}\Rightarrow D_{MD}(\widehat{i_{i}},\widehat{\tau_{i}})\le D_{MD}(\widehat{i_{i}},\widehat{\tau_{i}})$, where $i,j\in[1,N]$

\textbf{Proof}: Firstly, we expand $\widehat{R_{i}}$, $\widehat{R_{i}}$ and $R_{i}$ according to the algorithm.
\begin{eqnarray}
\widehat{R_{i}}&=&E(\widehat{C_{i}})\cdot E(\widehat{\tau_{i}}) \nonumber \\
&=& M_{1}\widehat{Q_{\tau}} \widehat{c_{i}}M_{2}M^{-1}_{2}\widehat{Q} \widehat{Q_{\tau}}M^{-1}_{1}\nonumber \\
&=& M_{1}\widehat{Q_{\tau}} \widehat{c_{i}}\widehat{Q} \widehat{Q_{\tau}}M^{-1}_{1} \nonumber\\
\overline{R_{i}}&=&E(\overline{C_{i}})\cdot E(\overline{\tau_{i}}) \nonumber \\
&=& M_{1}\overline{Q_{\tau}} \overline{c_{i}}M_{2}M^{-1}_{2}\overline{Q}\overline{Q_{\tau}}M^{-1}_{1}\nonumber \\
&=& M_{1}\overline{Q_{\tau}} \overline{c_{i}}\overline{Q} \overline{Q_{\tau}}M^{-1}_{1}\nonumber \\
{R_{i}}&=&E(\Psi^{'})\cdot E(\sum\nolimits_{i}) \nonumber \\
&=&M^{-1}_{4}\Psi^{'}M^{-1}_{3}M_3\sum\nolimits_{i}^{'}M_4   \nonumber \\
&=&M^{-1}_{4}\Psi^{'}\sum\nolimits_{i}^{'}M_4   \nonumber 
\end{eqnarray}
In this context, $\overline{C_{i}}$, $\widehat{C_{i}}$, $\widehat{Q}$, and $\overline{Q}$ represent diagonal matrices, while $\widehat{Q_{\tau}}$ and $\overline{Q_{\tau}}$, are lower triangular matrices with diagonal elements set to 1. Based on the properties of similarity matrices, it is evident that the trace of a similarity transformation matrix remains invariant, which means
\begin{eqnarray}
Tr(\widehat{R_{i}})&=&Tr(\widehat{c_{i}}\widehat{Q}) \nonumber \\             
Tr(\overline{R_{i}})&=&Tr(\overline{c_{i}}\overline{Q}) \nonumber \\  
Tr(R_{i})&=&Tr(\Psi^{'}\sum\nolimits_{i}^{'}) \nonumber 
\end{eqnarray}
Based on the above derivation, we can conclude that 
\begin{eqnarray}
Tr(CD_{\tau})&=&Tr(\widehat{R_{i}})+Tr(\overline{R_{i}})+Tr(R_{i})\nonumber 
\end{eqnarray}
\subsection{Security Analysis of PRISE}

 We claim that our proposed scheme can resist attack-\uppercase\expandafter{\romannumeral3} and that the PRISE scheme is secure under attack-\uppercase\expandafter{\romannumeral1} and attack-\uppercase\expandafter{\romannumeral2}. In our security proof design, the attacker may act as an authorized user capable of executing queries. Additionally, the cloud server may collude with the user to obtain as much private information from the image data.

\textbf{Theorem 1.} The PRISE scheme is capable of resisting attack-\uppercase\expandafter{\romannumeral3}, even in the presence of malicious authorized users colluding with the cloud server.

\textbf{Proof}:For attack-\uppercase\expandafter{\romannumeral3}, the attacker has access to the encrypted database and encrypted query tokens. If the attacker is the AU of the DO, they can obtain a set of samples from the database and know the corresponding plaintexts. The attacker has access to the encrypted database and encrypted query tokens. If the attacker is an authorized user of the data owner, they can obtain a set of samples from the database and know the corresponding plaintexts. For the attacker to obtain the result R from the encrypted database, they must acquire the keys M and N. The attacker may attempt to construct the following equation:

\begin{alignat*}{7}
&E(\widehat{\tau})=M^{-1}_{2}\widehat{Q} \widehat{Q_{\tau}}M^{-1}_{1}\nonumber \\             
&E(\overline{\tau})=M^{-1}_{2}\overline{Q} \overline{Q_{\tau}}M^{-1}_{1} \nonumber \\  
&E(\Psi^{'})=M^{-1}_{4}\Psi^{'}M^{-1}_{3} \nonumber\\
& U_1=M_{1}\times M^{-1}_{1} \nonumber \\
&U_2=M_{2}\times M^{-1}_{2} \nonumber \\
&U_3=M_{3}\times M^{-1}_{3} \nonumber \\
&U_4=M_{4}\times M^{-1}_{4} \nonumber 
\end{alignat*}

 Where $U_i$ is a  $(M+2)\times(M+2)$ identity matrix. In these equations, $M_{1}$, $M^{-1}_{1}$, $M_{2}$, $M^{-1}_{2}$, $M_{3}$, $M^{-1}_{3}$, $M_{4}$, $M^{-1}_{4}$, are eight random matrices. When an attacker attempts to solve for any of these matrices, they construct $2(M\times2)(M\times2)$ equations based on their knowledge. However, these equations are far fewer than the $(M\times2)^2$ unknowns in the equation. As a result, the attacker cannot recover the key from the known ciphertexts and corresponding plaintexts.

We depict the PRISE scheme utilizing the Real/Ideal simulation method \cite{link:19}\cite{link:35}. $(DataEnc, TokenGen, Query)$ components of a privacy-preserving searchable symmetric encryption method that is based on image data. Assuming $\mathcal{A}$ represent the challenger, $\mathcal{S}$ represent the simulator, $k$ represent the security parameter, and $\mathcal{L}$ is secure against adaptive CQA-secure.

$Real$ $Environment$. There exists a stateful polynomial-time adversary $\mathcal{A}$  and a challenger $\mathcal{C}$ , and they interact with each other to perform the following operations.

\begin{itemize}
	\item Initialization phase: $\mathcal{A}$ randomly construct a dataset $\mathcal{D}$ consisting of $N$ images. Then, $\mathcal{A}$ perform feature extraction using the MPCA algorithm. Finally, output the feature and covariance matrix $\sum_{A}$, send the results to $\mathcal{C}$.
	\par
	\item Encryption phase: $\mathcal{C}$ first generates the key $sk_1=\lbrace M_{1},M_{2} \rbrace$, and then executes the DataEnc($\mathcal{D}$,$\sum_{A}$,$sk_1$) algorithm, the algorithm outputs the encrypted database $\lbrace$ $E(s_{1}), E(s_{2})$,..., $E(s_{N})\rbrace$.
	\par
	\item Query phase:$\mathcal{A}$ selects a set of queried $\mathcal{q}$ and sends them to the $\mathcal{C}$. The challenger $\mathcal{C}$ executes the TokenGen($\mathcal{q},sk_{1}$) algorithm and eventually returns the encrypted tokens $q_{\tau}$ to 
	$\mathcal{A}$.
   \par
    \item Challenge phase: $\mathcal{C}$ receives a query token $q_{\tau}$ from $\mathcal{A}$. First, $\mathcal{C}$ determines the class to which the query token $E(\tau)$ belongs through these computation $ \widehat{R_{i}}=E(\widehat{C_{i}})\cdot E(\widehat{\tau_{i}}) $, $ \overline{R_{i}}=E(\overline{C_{i}})\cdot E(\overline{\tau_{i}})$,  $CD_{\tau}=Tr(\widehat{R_{i}})+Tr(\overline{R_{i}})+Tr({R_{i}})$. $\mathcal{C}$ returns the results $R_{\tau}$ containing the token $q_{\tau}$ to $\mathcal{A}$. 
\end{itemize}

$Ideal$ $Environment$. The ideal environment of PRISE consists of a stateful adversary  $\mathcal{A}$ and a simulator $\mathcal{S}$ equipped with a leakage function.

\begin{itemize}
	\item Initialization phase: $\mathcal{A^{'}}$ begins by constructing a database $\mathcal{D^{'}}$ comprising $N$ image samples and calculating the corresponding covariance matrix $\sum^{'}_{A}$. $\mathcal{A^{'}}$ sends the dataset and the covariance matrix $\sum^{'}_{A}$ to $\mathcal{S}$.
	\par
	\item Encryption phase: $\mathcal{S}$ first creates the key $sk^{'}_1=$$\lbrace M^{'}_{1},$$M^{'}_{2}\rbrace$, and then executes the DataEnc($\mathcal{D^{'}}$,$\sum^{'}_{A}$,$sk^{'}_1$) algorithm, outputs the encrypted database $\lbrace$ $E^{'}(s_{1}), E^{'}(s_{2})$,..., $E^{'}(s_{N})\rbrace$.
	\par
	\item Query phase: $\mathcal{A^{'}}$ elects a group of queried $\mathcal{q^{'}}$ and sends them to the $\mathcal{S}$. The challenger $\mathcal{A^{'}}$ executes the TokenGen($\mathcal{q^{'}},sk^{'}_{1}$) algorithm, completes the process by sending the encrypted tokens $q^{'}_{\tau}$ to $\mathcal{A^{'}}$.
	\par
	\item Challenge phase: $\mathcal{A^{'}}$ sends $q^{'}_{\tau}$ to $\mathcal{S}$. Initially, $\mathcal{S}$ determines the cluster to which the query token $q^{'}_{\tau}$ belongs through these computation $ \widehat{R^{'}_{i}}=E(\widehat{C^{'}_{i}})\cdot E^{'}(\widehat{\tau_{i}}) $, $ \overline{R^{'}_{i}}=E^{'}(\overline{C_{i}})\cdot E^{'}(\overline{\tau_{i}})$,  $CD^{'}_{\tau}=Tr(\widehat{R^{'}_{i}})+Tr(\overline{R^{'}_{i}})+Tr({R^{'}_{i}})$. $\mathcal{S}$ returns the results $R^{'}_{\tau}$ to $\mathcal{A^{'}}$. 
\end{itemize}

We provide a security proof for the proposed PRISE scheme based on $\mathcal{A}$ in both real and ideal environments. 

\textbf{Theorem 2.} The PRISE scheme is indistinguishable under the leakage function $\mathcal{L}$ in the KPA model for any PPT adversary. 

\textbf{Proof}: We will provide evidence that points to the existence of a PPT simulator $\mathcal{S}$ and a probabilistic $\mathcal{A}$.
If $\mathcal{A}$ cannot distinguish whether the result comes from the ideal environment or the real environment, we prove that the scheme is secure.
 First, $\mathcal{S}$ simulates the encryption algorithm to encrypt image data ${i_{q}}^{'}$ by randomly selecting iamge data. $\mathcal{S}$ randomly selects image query vectors $Q_{\tau}^{'}$. $\mathcal{A}$ is unable to discriminate between $({i_{q}}^{'}, Q_{\tau}^{'})$ and $(i_{q}, Q_{\tau})$. $\mathcal{A}$ performs a query, let $\mathcal{S}$ simulates a query token $\tau_{q}$. $\mathcal{A}$ obtains the token $\tau_{q}$. $\mathcal{A}$ is unable to distinguish between authentic token $\tau_{q}$ and additional tokens $Q_{\tau}^{'}$ for subsequent queries. 

 The encryption mechanism uses four random invertible matrices $M_1, M_2, M_3, M_4$. To recover private data  $I_p$, the attacker must solve the following system derived from the encryption equations:
\begin{alignat*}{7}
&E(\widehat{\tau})=M^{-1}_{2}\times \widehat{Q} \times \widehat{Q_{\tau}}\times M^{-1}_{1}\nonumber \\             
&E(\overline{\tau})=M^{-1}_{2}\times \overline{Q} \times \overline{Q_{\tau}}\times M^{-1}_{1} \nonumber \\  
&E(\Psi^{'})=M^{-1}_{4}\times \Psi^{'}M^{-1}_{3} \nonumber\\
& U_1=M_{1}\times M^{-1}_{1} \nonumber \\
&U_2=M_{2}\times M^{-1}_{2} \nonumber \\
&U_3=M_{3}\times M^{-1}_{3} \nonumber \\
&U_4=M_{4}\times M^{-1}_{4} \nonumber 
\end{alignat*}
where $U_i$ is a $(n+2)\times (n+2)$ identity matrix. In these equations, $M_{1}, M_{2}, M_{3}, M_{4}, M^{-1}_{1}, M^{-1}_{2}, M^{-1}_{3}$and $M^{-1}_{4}$ are $8(n+2)*(n+2)$ random unknown matrices, $\widehat{Q}$ and $\overline{Q}$ are  $2(n+2)*(n+2)$ random lower triangular matrices.
To solve $M_1$, attacker need build $(n + 2)\times(n + 2)$ independent equations based on his/her knowledge, which is not enough for the $(n + 2)\times(n + 2)$ unknowns in the equations. For $ M_{2}, M_{3}, M_{4}, M^{-1}_{1}, M^{-1}_{2}, M^{-1}_{3}$and $M^{-1}_{4}$, similar to $M_1$, unknowns in $7(n + 1)^2$ independent equations also cannot be solved.  Therefore, the attackers cannot directly recovery secret keys  $M_{1}, M_{2}, M_{3}, M_{4}, M^{-1}_{1}, M^{-1}_{2}, M^{-1}_{3}$and $M^{-1}_{4}$ $\widehat{Q}$ and $\overline{Q}$ with these ciphertexts and the corresponding plaintexts.

When the attacker himself is a legitimate client, he can submit an identification request and collude with the cloud server. Consequently, in addition to the knowledge he already possesses, he also gains extra information $Q_{\tau}$. The attacker can construct  additional equations using the encrypted request as follows:
\begin{alignat*}{2}
& \widehat{R_{i}}=E(\widehat{C_{i}})\cdot E(\widehat{\tau_{i}}) \\
& \overline{R_{i}}=E(\overline{C_{i}})\cdot E(\overline{\tau_{i}}) 
\end{alignat*}

Although the attacker knows $Q_{\tau}$, it remains obfuscated by random numbers in  $E(\widehat{C_{i}}) $. Therefore, these newly introduced equations also introduce $2(n + 2)\times(n + 2)$ new unknowns, providing no additional advantage for the attacker in recovering private data. For an attacker attempting to recover the candidate data $\widehat{R_{i}}$, it is necessary to solve $M_{1}, M_{2}, M_{3}$ and  $M_{4}$. However, the attacker only has $4(n + 2)\times(n + 2)$  equations for $M^{-1}_{1}, M^{-1}_{2}, M^{-1}_{3}$, $M^{-1}_{4}$ $\widehat{Q}$ and $\overline{Q}$ , whereas the total number of unknowns is $6(n + 2)\times(n + 2)$, which exceeds the number of available equations. Therefore, based on the above analysis, a attack-\uppercase\expandafter{\romannumeral3} attacker cannot acquire enough knowledge to access the private data in the image feature database and user data.

We declare that the scheme is secure if the attacker $\mathcal{A}$ operates in PPT time.We claim that the scheme is secure under the security parameter $\lambda$ while attacker $\mathcal{A}$ conducts in PPT time. $|Pr[Real_{PRISE,\mathcal{A}}(\lambda)]-Pr[Ideal_{PRISE,\mathcal{A}}(\lambda)]|\leq negl(\lambda)$.

\section{Implementation and Evaluation}
\label{sec:perf}

The rapid development of smart vehicles has created significant potential for improving traffic safety, efficiency, and comfort, but it has also raised concerns about the privacy and security of onboard data. In this context, image searchable encryption schemes have emerged as essential tools for protecting the privacy of onboard image data.

In this section, we conduct experiments to analyze the proposed solution and compare it with other approaches. In terms of theoretical analysis, we focus on various metrics, including the computational complexity, communication complexity, communication overhead, computational cost, storage requirements, error rates, and other relevant metrics of the proposed solution. 

To evaluate the feasibility of our proposed approach, we implemented it using Python 3.7 and conducted all validation experiments within the PyCharm (Professional edition). The tests were run on a machine equipped with an Intel(R) Core(TM) AMD R9 3900X CPU, 16 GB of RAM, and running the 64-bit version of Windows 10. We used four datasets—INRIA Holidays, Mirflickr, GTSRB, and CIFAR-100 to assess the performance of image similarity search. Our experiments focused on key aspects such as initial construction time, search time, storage overhead, and accuracy.

The INRIA Holidays dataset consists of 1,491 high-quality images, grouped into 500 sets. Each set typically includes one reference image and several related transformed images (e.g., rotated, cropped, or with varying illumination). The Mirflickr dataset comprises 25,000 images collected from the Flickr website, accompanied by user-generated tags, titles, and descriptions. These images cover a wide range of categories, including nature, architecture, humans, and animals. The GTSRB (German Traffic Sign Recognition Benchmark) dataset contains 51,839 traffic sign images categorized into 43 classes. These images were captured in real-world road environments in Germany, exhibiting diverse lighting, weather, and perspective conditions. The CIFAR-100 dataset is a well-known image classification dataset provided by the University of Montreal and is widely used in deep learning and image recognition research. It includes 60,000 color images, divided into 100 classes, with 600 images per class.

\subsection{Complexity Analysis}

This section provides an in-depth analysis of the computation complexity and communication complexity of the searchable encryption scheme based on images for intelligent vehicle. Firstly, we focus on the computational complexity of the scheme.

We analyze the computational complexity from three phases: the encryption phase, token generation phase, and query phase. The computational complexity of an n × n-dimensional matrix is $\mathcal{O}(n^2)$, so we assume the time complexity of matrix multiplication is $\mathcal{O}(n^3)$. In the PRISE scheme, the computational complexity of the encryption phase is $\mathcal{O}(N(n+2)^3)$, the computational complexity of the token generation phase is $\mathcal{O}(C(n+2)^3)$, and the computational complexity of the search phase is $\mathcal{O}(C_i(n+2)^2)$. Here, $N$ represents the number of images, $n$ denotes the dimensionality of the vectors, and $C$ represents the number of indices in a particular class.

Next, we analyze the communication complexity of the proposed scheme. Data from intelligent vehicle needs to communicate with cloud servers to facilitate functionalities such as data retrieval and recommendations. Therefore, the communication overhead associated with the encryption and search processes of image data has a significant impact on system performance and user experience. We analyze the communication complexity from the encryption phase, token generation phase, and query phase. We assume the communication complexity of matrices is  $\mathcal{O}(n^2)$. In the PRISE scheme, the communication complexity of the encryption phase is $\mathcal{O}(N(n+2)^2)$, the communication complexity of the token generation phase is $\mathcal{O}(C_i(n+2)^2)$, and the communication complexity of the search phase is $\mathcal{O}(k)$, where $k$ is the number of similar images returned. 

It is worth noting that although the complexities of computation and communication increase linearly with the size of the database, these are one-time costs that do not affect the real-time performance of image recognition processes and user experience. We offload storage and computation to the cloud, taking advantage of its high storage and computational capabilities.

\subsection{MPCA Based on FCM-MD}

To validate the effectiveness of the Mahalanobis Distance-Based Fuzzy Multilinear Principal Component Analysis scheme in image retrieval, we will conduct experiments to assess the proposed approach from multiple perspectives, including the performance of dimensionality reduction, the number of principal components, noise resistance, and parameter sensitivity.

The data is subjected to dimensionality reduction using MPCA, resulting in reduced feature sizes significantly smaller than the original features, demonstrating effective dimensionality reduction. Meanwhile, the amount of information retained in the reduced features can be evaluated using metrics such as the explained variance ratio or the cumulative variance contribution rate. Assume we have a dataset, and after applying MPCA for dimensionality reduction, we obtain three principal components with eigenvalues $a_{1}$$=$4.5, $a_{2}$$=$2.3, $a_{3}$$=$1.2. The steps to calculate the variance contribution rate are as follows:total variance:$a_1+a_2+a_3=8$, total variance: $a_1+a_2+a_3=8$, first principal component:56.25\%, first two principal components:$56.25\%+28.75\%=85\%$, first three principal components:$56.25\%+28.75\%+15\%=100\%$. 

{We randomly selected 100 images from multiple datasets to perform MPCA feature extraction. Based on the calculation of the above variance contribution rate, we conducted 20 experiments to perform MPCA on RGB three-channel images to evaluate the variance contribution rate. Finally, we calculated the average of the experimental results. The cumulative contribution rates of the first three principal components are $78\%$, $95\%$, and $100\%$, respectively.
	
\begin{table}[h]
	\centering
	\caption{The recall rate under different noise levels}
	\centering
	\begin{tabular}{c|c|c|c|c|c|c}
		\hline
		$\sigma$ &0& 0.01 &  0.05& 0.1 &0.2& 0.5\\
		\hline
		$\gamma$ & 87.8\%  & 87.8\%  & 87.7\%&87.5\%&87.3\% &86.8\%\\
		
		\hline
	\end{tabular}
	\label{table:resistance}
\end{table}	
	
We observe whether Mahalanobis distance remains stable in fuzzy C-means clustering by adding Gaussian noise. To validate the stability and noise robustness of the PRISE, we conducted evaluations by adding Gaussian noise with different standard deviations $\sigma$ to the dataset. The experimental results demonstrate that the proposed method maintains high performance under various noise conditions, showcasing excellent noise robustness. The results are presented in Table~\ref{table:resistance}, where $\sigma$ represents the standard deviation and $\gamma$ represents the recall rate. According to Table~\ref{table:resistance}, the recall rate of the proposed scheme remains relatively stable under different levels of Gaussian noise $\gamma$.

By adjusting the number of clusters and the fuzzy coefficient parameters in the Fuzzy C-Means (FCM) classification, we examine their impact on retrieval time overhead and retrieval accuracy to determine the optimal settings. In analyzing the effect of the fuzzy coefficient for the proposed scheme, we fix the number of clusters to 500 to observe how different fuzzy coefficient values influence query time and accuracy. The experimental results are shown in Table~\ref{table:cluster} and Table~\ref{table:fuzzy}. According to Table~\ref{table:cluster}, it can be observed that as the number of categories increases, the query time increases, while the recall rate relatively improves and the error rate decreases. From Table~\ref{table:fuzzy}, it can be seen that under different fuzzy coefficients, the query time remains relatively stable, and both the recall rate and error rate remain relatively stable, indicating that the fuzzy coefficient has a relatively small impact on the proposed scheme.
\begin{table}[h]
	\centering
	\caption{The analysis under different numbers of clusters}
	\centering
	\begin{tabular}{c|c|c|c|c|c}
		\hline
		Cluster &400& 450 &  500&550 &600\\
		\hline
		Query time(s)& 1.05  & 1.19  & 1.32 &1.45 &1.56  \\
		\hline
		Recall rate& 86.2\%  & 87.1\%  & 87.8\%&88.0\%&88.0\% \\
		\hline
		Error rate& 12.3\%  & 11.6\%  & 11.2\%&11.2\%&11.2\% \\
		\hline
	\end{tabular}
	\label{table:cluster}
\end{table}

\begin{table}[h]
	\centering
	\caption{The analysis under different fuzzy coefficient}
	\centering
	\begin{tabular}{c|c|c|c|c|c}
		\hline
		Fuzzy coefficient &1.5& 1.8 &  2.0&2.3 &2.5\\
		\hline
		Query time(s)& 1.31  & 1.32  & 1.32 &1.33 &1.33  \\
		\hline
		Recall rate& 87.4\%  & 87.5\%  & 87.8\%&88.7\%&87.5\% \\
		\hline
		Error rate& 11.4\%  & 11.2\%  & 11.1\%&11.2\%&11.4\% \\
		\hline
	\end{tabular}
	\label{table:fuzzy}
\end{table}

We also conducted experiments to evaluate the impact of cumulative contribution rate on the recall rate of the proposed method. The results are shown in Table~\ref{table:Contributionpca}. From the experimental results, it can be observed that the number of principal components extracted by the MPCA method ranges from 87 to 97, with only slight differences in recall rate. However, as the number of principal components increases, the recall rate also improves.
\begin{table}[h]
	\centering
	\caption{The analysis under different cumulative contribution rate}
	\centering
	\begin{tabular}{c|c|c|c|c|c}
		\hline
		Contribution rate& 87\%  & 90\%  & 93\%&95\%&97\% \\
		\hline
		Recall rate& 85.1\%  & 85.6\%  & 86.9\%&87.8\%&87.8\% \\
		\hline
	\end{tabular}
	\label{table:Contributionpca}
\end{table}

\subsection{System Initialization Cost}
In the init phase, we primarily assess the time required to generate a secure index. The overhead associated with generating a secure index consists of two main components: the time required for extracting image features, and the time needed to encrypt these features. The validation of the proposed solution is conducted using two real-world datasets.
\begin{figure}[h]
	\centering
	\includegraphics[width=3.0in]{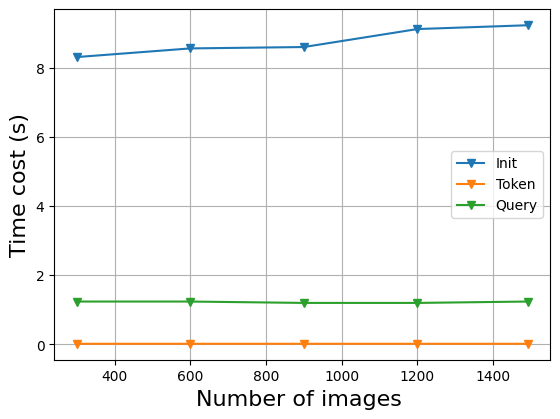}
	\caption{The time overhead of  the PRISE scheme on the INRIA dataset}
	\label{fig:inria}
\end{figure}
Fig.\ref{fig:inria} illustrates the validation experiments conducted on the INRIA dataset, which consists of a total of 1491 images. We evaluate the time required for generating an encrypted index. The horizontal axis in Fig.\ref{fig:inria} represents the number of images, while the vertical axis depicts the time taken for encrypting the features. From the Fig.\ref{fig:inria}, it can be observed that the time required to generate the encrypted index increases linearly with the number of images.

\begin{figure}[h]
	\centering
	\includegraphics[width=3.0in]{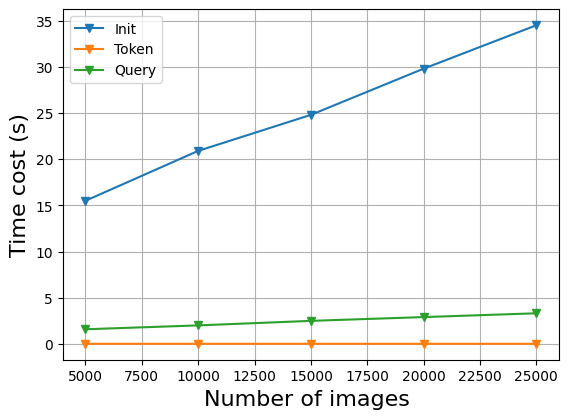}
	\caption{The time overhead of the PRISE scheme on the Mirflickr dataset.}
	\label{fig:mirflickr}
\end{figure}
Fig.\ref{fig:mirflickr} shows the validation experiments conducted on the Mirflickr dataset, which consists of a total of 25,000 images. We evaluate the time required to generating an encrypted index. The horizontal axis in Fig.\ref{fig:mirflickr} represents the number of images, while the vertical axis illustrates the time taken for encrypting the features. It is evident from the Fig.\ref{fig:mirflickr} that the time required to generate an encrypted index increases linearly with the number of images.

\subsection{Token Generation Cost}
When authorized users query data on the cloud, it is necessary to generate an encrypted token to prevent information leakage. In the token generation phase, the time overhead primarily consists of two parts: the construction time of the matrix and the time required to generate the encrypted vector using matrix encryption. The total time for token generation is the sum of these two components.

Fig.\ref{fig:inria} depicts validation experiments conducted on the INRIA dataset. The horizontal axis in Fig.\ref{fig:inria} represents the number of images, while the vertical axis illustrates the token generation time. Fig.\ref{fig:mirflickr} displays validation experiments conducted on the Mirflickr dataset. From the Fig.\ref{fig:inria} and Fig.\ref{fig:mirflickr}, it can be observed that the time remains constant. The token generation on the INRIA dataset takes approximately 18.01ms, while on the Mirflickr dataset, it requires around 18.07ms.

\subsection{Retrieval Cost}

 In this section, we analyze the retrieval performance of PRISE and compare it with state-of-the-art related works \cite{link:39}, \cite{link:40} and \cite{link:42}. The VFirm scheme in \cite{link:39} proposed verifiable encrypted image retrieval in multi-owner multi-user settings, focusing on retrieval accuracy, access control, and verifiability.  It employs SE and fine-grained access control, which are also relevant to encrypted image similarity retrieval. The TAMMIE scheme in \cite{link:40} explored accurate and privacy-preserving retrieval of outsourced medical images, proposing a method that combines deep learning-based image feature extraction, locality-sensitive hashing, and SE. Its objective aligned closely with ours—enhancing retrieval efficiency and accuracy while ensuring privacy protection.  Abdul et al. \cite{link:42} proposed novel encryption and indexing schemes to securely store sensitive images on cloud servers and facilitate their secure retrieval by authorized users. These three schemes, along with the scheme proposed in this paper, are all encrypted image retrieval schemes, so they are suitable as benchmarks for comparative analysis.

When the cloud server receives a token from an authorized user, the cloud server queries the data stored by the data owner in the cloud. This phase involves two primary time overheads: one for identifying the class of the token and another for calculating the top-k results within that class. Four methods for verifying query schemes were compared. The vector scheme utilizes encrypted vectors for search operations. This paper proposes the MPCA scheme, which employs the MPCA method for feature extraction, and employs matrix encryption to protect feature privacy. The PRISE scheme extends the MPCA scheme through the application of the fuzzy C-means method for data classification. TAMMIE is a scheme proposed in \cite{link:40}, while VFIM is introduced in \cite{link:39}. Fig \ref{fig:inqu} illustrates that the PRISE scheme outperforms both the TAMMIE and VFIM schemes. The PRISE scheme outperforms both the vector and MPCA schemes when compared to non-classification-based approaches. Fig \ref{fig:miqu} illustrates that the query time for the vector and MPCA schemes increases linearly with the number of images, whereas the PRISE, TAMMIE, and VFIM schemes exhibit relatively stable query times. In conclusion, the PRISE scheme presented in this paper is feasible.

\begin{figure}[h]
	\centering
	\includegraphics[width=3.0in]{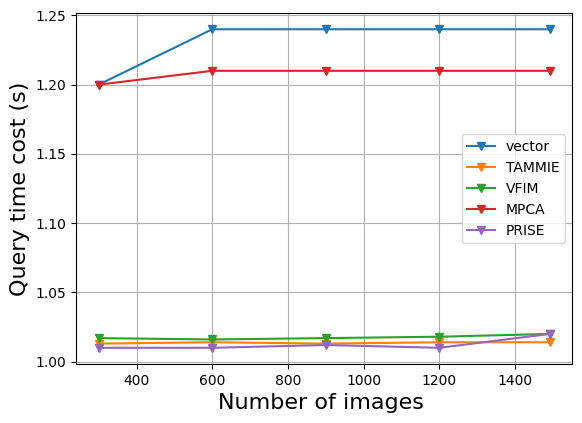}
	\caption{The time overhead of query on the INRIA dataset.}
	\label{fig:inqu}
\end{figure}

\begin{figure}[h]
	\centering
	\includegraphics[width=3.0in]{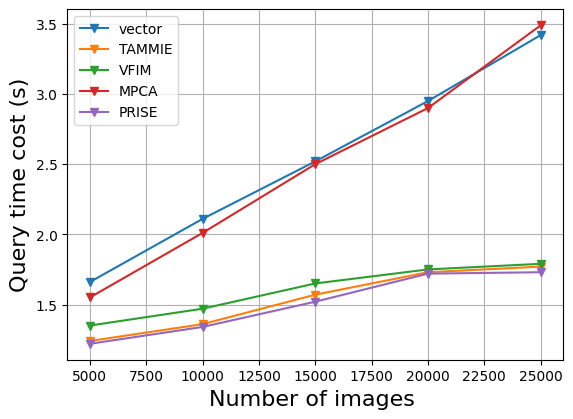}
	\caption{The time overhead of query on the Mirflickr dataset.}
	\label{fig:miqu}
\end{figure}

We compared two different image searchable encryption schemes: VFRM \cite{link:39}, and TAMMIE \cite{link:40}. For each scheme, we evaluated its performance in terms of retrieval accuracy, and query time overhead. We adopted the same evaluation criteria for each scheme to ensure fair comparison. For retrieval accuracy, we used accuracy  and recall as evaluation metrics, measuring the performance by comparing the matching degree between the retrieval results returned by each scheme and the ground truth results. The comparative experimental results of the three schemes are shown in Table~\ref{table:accuracy} and Table~\ref{table:recall}. By comparing the experimental results, we found that PRISE performed the best in terms of retrieval accuracy, with both precision and recall higher than VFRM, TAMMIE and LCLD.

\begin{table}[h]
	\centering
 \caption{The comparison of  accuracy under different schemes}
	\centering
	\begin{tabular}{c|c|c|c|c}
		\hline
		\diagbox{Dataset}{Scheme}	 & VFRM\cite{link:39} &  TAMMIE \cite{link:40}&LCLD \cite{link:42}& PRISE\\
		\hline
		INRIA & 0.70  & 0.75 & 0.70&0.82\\
		\hline
		mirflickr & 0.72  & 0.74 &0.68& 0.77 \\
		\hline
	\end{tabular}
	\label{table:accuracy}
\end{table}

\begin{table}[h]
	\centering
	\caption{The comparison of recall rates under different schemes}
	\centering
	\begin{tabular}{c|c|c|c|c}
		\hline
		\diagbox{Dataset}{Scheme}	 & VFRM\cite{link:39} &  TAMMIE \cite{link:40}&LCLD \cite{link:42}& PRISE\\
		\hline
		INRIA & 0.62  & 0.64 &0.68& 0.87\\
		\hline
		mirflickr & 0.60  & 0.62 & 0.64&0.83 \\
		\hline
	\end{tabular}
	\label{table:recall}
\end{table}

We validate the robustness of the proposed method by extracting features using MPCA on different datasets and performing classification using FCM clustering based on Mahalanobis distance. The effectiveness of the method is evaluated by testing its recall and error rates on datasets from various scenarios. The experimental results are presented in Table~\ref{table:cmpare1}. From the results, it can be concluded that the proposed method demonstrates relatively stable recall rates, with an average recall rate consistently maintained at 0.86 and an error rate around 0.1.

\begin{table}[h]
	\centering
	\caption{The comparison of recall rates under different datasets}
	\centering
	\begin{tabular}{c|c|c}
		\hline
		\diagbox{Dataset}{Result}	 & recall rate &  error rate\\
		\hline
		INRIA & 0.87.  & 0.11 \\
		\hline
		mirflickr & 0.83  & 0.12  \\
		\hline
		GTSIRB & 0.89  & 0.09  \\
		\hline
		CIFAR-100 & 0.85  & 0.13  \\
		\hline
	\end{tabular}
	\label{table:cmpare1}
\end{table}

We validated the recall and error rates of different feature extraction methods on the same dataset. As shown in Table~\ref{table:compare2}, the proposed method outperforms both LDA and PCA in terms of recall and error rate.

\begin{table}[h]
	\centering
	\caption{The comparison of recall rates under different feature extraction methods.}
	\centering
	\begin{tabular}{c|c|c|c}
		\hline
		\diagbox{Result}{Scheme} & LDA &  PCA& MPCA\\
		\hline
		recall rate & 0.79.  & 0.78& 81.3 \\
		\hline
		 error rate & 0.20  & 0.22& 0.18  \\
			\hline
	\end{tabular}
	\label{table:compare2}
\end{table}

\section{Future Direction and Conclusion}
\label{sec:fut}

This paper proposes a novel and privacy-preserving image retrieval scheme. The proposed approach incorporates an innovative feature extraction method to extract features from images, utilizing matrix encryption to secure these features. To boost search efficiency, a FCM clustering algorithm based on Mahalanobis distance is introduced to expedite server search operations, thereby enhancing user experience. Furthermore, the paper provides a comprehensive security analysis and experimental validation of the proposed solution. The experimental results demonstrate the effectiveness of the encrypted searchable scheme, based on Mahalanobis distance and FCM, in safeguarding user image data privacy from disclosure.

\bibliographystyle{abbrvnat}
\bibliography{mybib}
\begin{IEEEbiography}[{\includegraphics[width=1in,height=1.25in,clip,keepaspectratio]{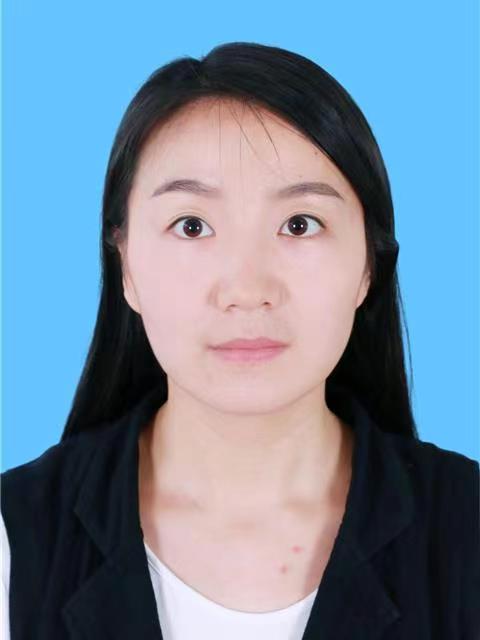}}]
    {Xuemei Fu} is currently pursuing the Ph.D. degree with the School of Information and Communication Engineering, Hainan University, Haikou, China. She received the master’s degree in Computer Science and Technology form Guangxi Normal University, Guilin, China, in 2021.
     Her current research interests include searchable encryption and cloud security.
\end{IEEEbiography}



\begin{IEEEbiography}[{\includegraphics[width=1in,height=1.25in,clip,keepaspectratio]{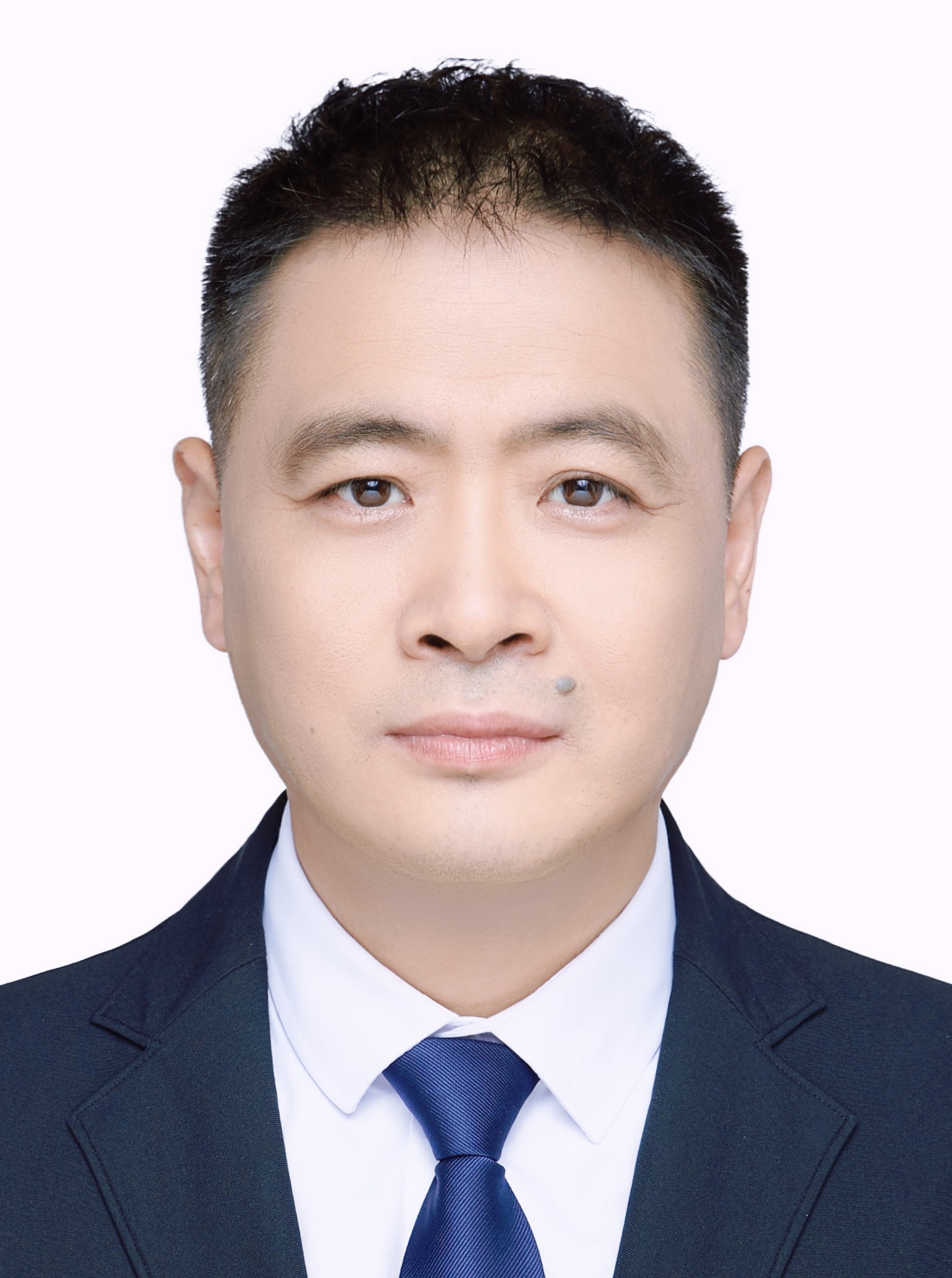}}]
	{Laurence T. Yang} (Fellow, IEEE) received the B.E. degree in Computer Science and Technology and B.Sc. degree in Applied Physics both from Tsinghua University, Beijing, China, in 1992, and the Ph.D. degree in Computer Science from the University of Victoria, Victoria, BC, Canada, in 2006. 

	He is a Professor in the School of Computer Science and Artificial Intelligence, Zhengzhou University, Zhengzhou, China, also with the School of Computer Science and Technology, Huazhong University of Science and Technology, Wuhan, China and with the Department of Computer Science, St. Francis Xavier University, Antigonish, NS B2G 2W5, Canada. His research interests include parallel and distributed computing, embedded and ubiquitous computing, and big data. His research has been supported by the National Sciences and Engineering Research Council and the Foundation for Innovation of Canada, as well as  by the National Natural Science Foundation of China and Ministry of Science and Technology of China.
	
\end{IEEEbiography}

\begin{IEEEbiography}[{\includegraphics[width=1in,height=1.25in,clip,keepaspectratio]{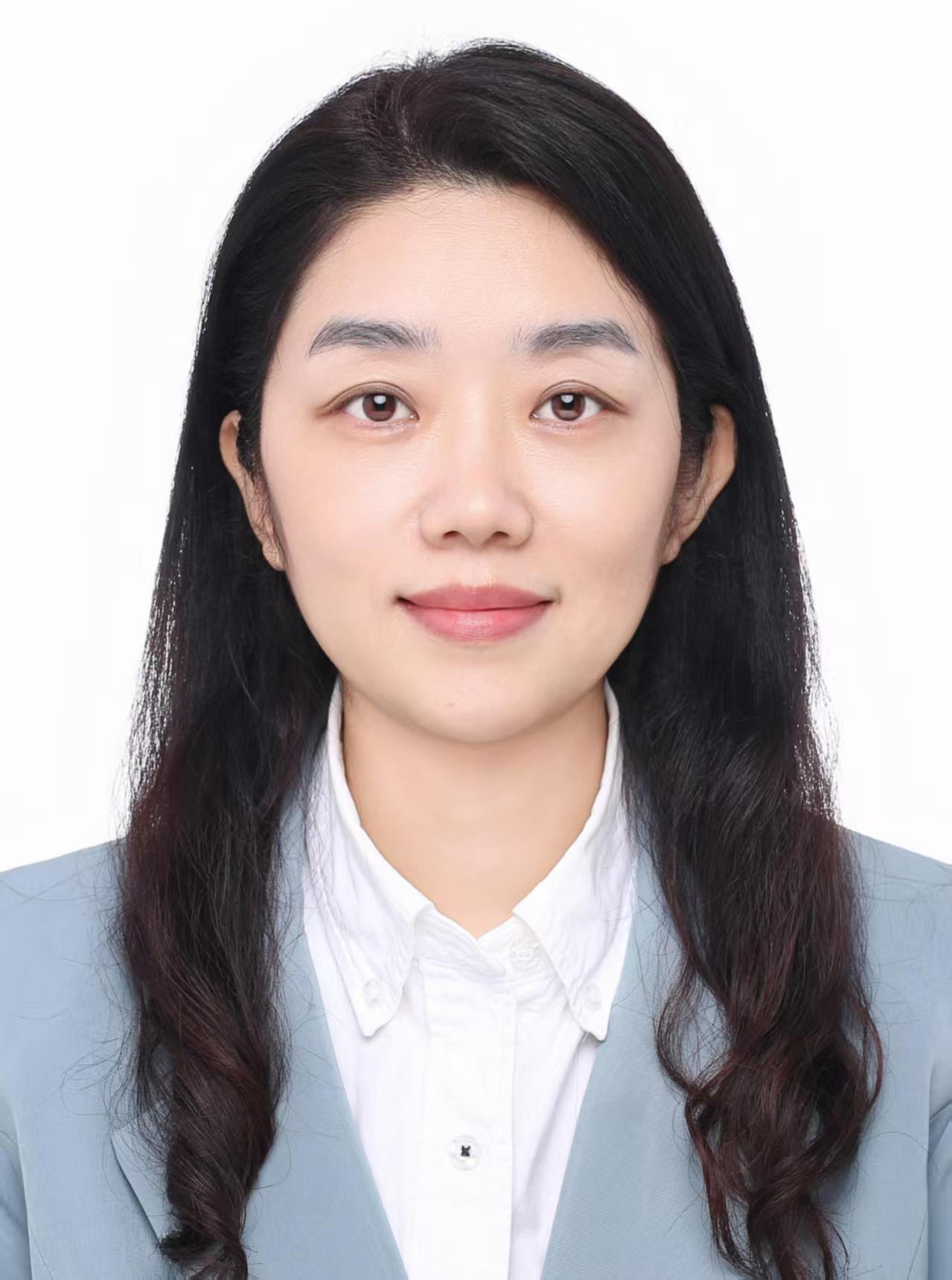}}]
	{Na Song} received the MEng degree in signal and information processing from Shenzhen University, China in 2010. She is currently a lecturer at School of 	Mechanical, Electrical and Information Engineering, Putian University, China. Her research interests include deep learning, multi-modal learning, and explainable neural networks.
\end{IEEEbiography}

\begin{IEEEbiography}[{\includegraphics[width=1in,height=1.25in,clip,keepaspectratio]{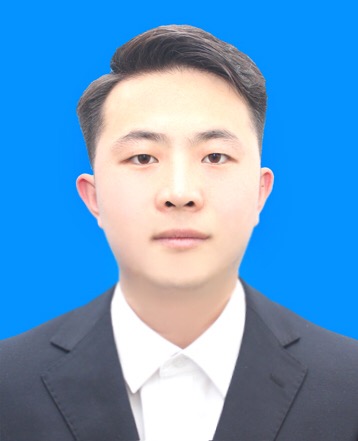}}]
	{Jinxiong Gao } received the B.S. degree in electronic information engineering from Inner Mongolia University, China, in 2018, and the M.S. degree in transportation information engineering from Inner Mongolia University, China, in 2021. He is currently pursuing the Ph.D. degree in the School of Information and Communication Engineering, Hainan University, China.
	His research interests include computer vision and deep learning.
\end{IEEEbiography}

\begin{IEEEbiography}[{\includegraphics[width=1in,height=1.25in,clip,keepaspectratio]{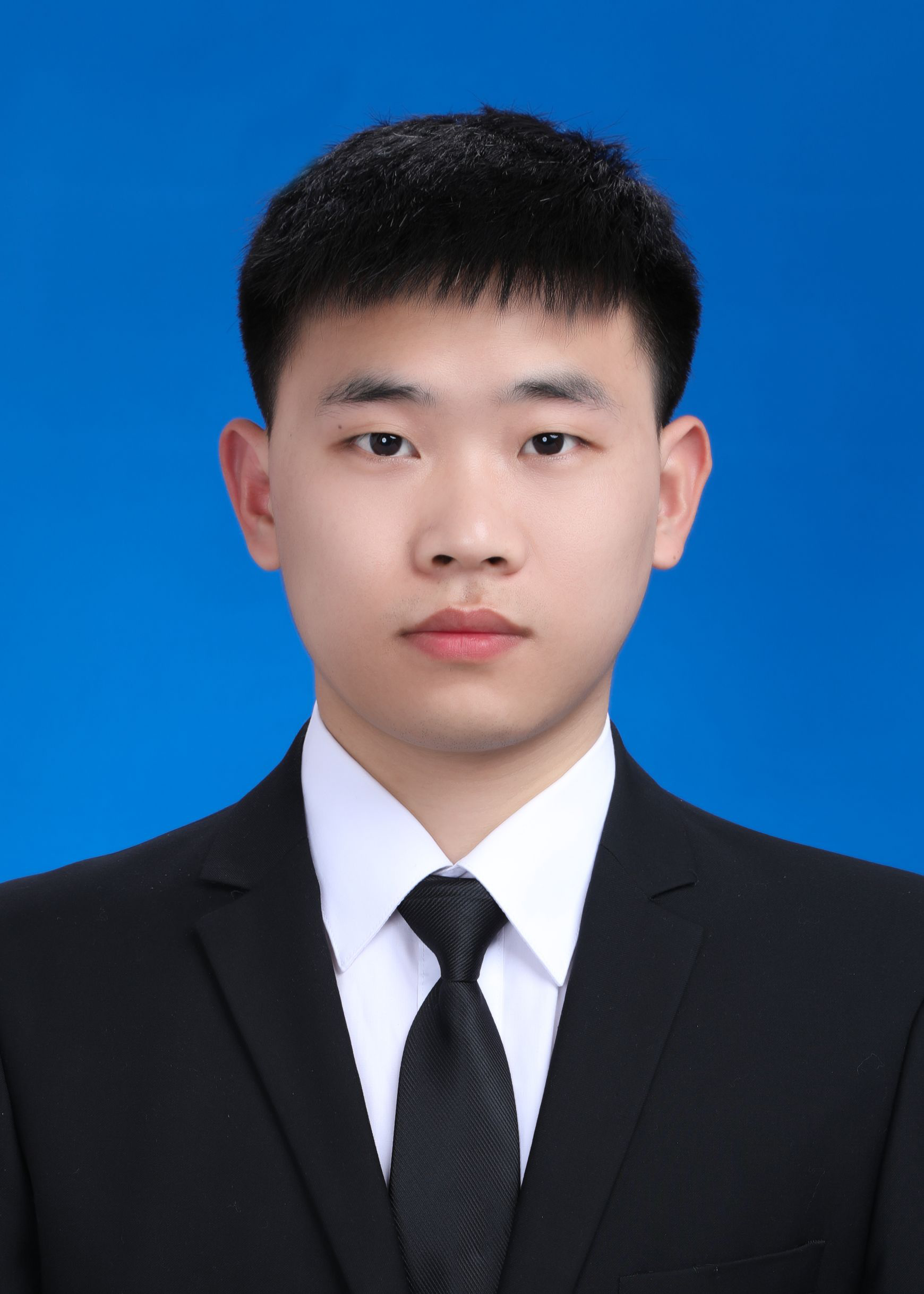}}]
	{Zhixing Lu} rceived the M.S. degree in the School of Infomation Management, Jiangxi University of Finance and Economics, China. He is currently pursuing the Ph.D. degree with the Faculty of Computer Science and Information Technology, Universiti Putra Malaysia. His research interests include Tensor Computing, Behavior Learning, Big DataAnalysis, and Privacy Protection. 
\end{IEEEbiography}





\end{document}